\documentclass[10pt,onecolumn]{IEEEtran}
\usepackage{amsthm}
\usepackage{times,amssymb,amsmath,amsfonts,float,nicefrac,color,bbm,mathrsfs,caption,float}
\usepackage{algorithm,enumerate,multirow,caption,tikz,graphicx}
\usepackage[mathscr]{eucal}
\usepackage{sidecap}
\usepackage{algpseudocode}
\usepackage{verbatim}
\usepackage{epstopdf}
\usepackage{textcomp}
\usetikzlibrary{shapes,arrows}
\usepackage{stmaryrd}
\usepackage{mathabx}
\usepackage[noadjust]{cite}
\usepackage{booktabs}
\usepackage[top=1in,bottom=1in,left=1in,right=1in]{geometry}
\allowdisplaybreaks

\newtheorem{theorem}{Theorem}[section]

\newtheorem{lemma}[theorem]{Lemma}

\newtheorem{definition}[theorem]{Definition}
\newtheorem{example}[theorem]{Example}
\newtheorem{corollary}[theorem]{Corollary}

\newtheorem{cnstr}{Construction}

\newtheorem{remark}[theorem]{Remark}

\newcommand{\RN}[1]{%
  \textup{\expandafter{\romannumeral#1}}%
}

\newcommand\remove[1]{}
\newcommand{\nc}{\newcommand}

\allowdisplaybreaks
\nc\bfa{{\boldsymbol a}}\nc\bfA{{\boldsymbol A}}\nc\cA{{\mathcal A}}\nc\sA{{\mathscr A}}
\nc\bfb{{\boldsymbol b}}\nc\bfB{{\boldsymbol B}}\nc\cB{{\mathcal B}}\nc\sB{{\mathscr B}}
\nc\bfc{{\boldsymbol c}}\nc\bfC{{\boldsymbol C}}\nc\cC{{\mathcal C}}\nc\sC{{\mathscr C}}
\nc\bfd{{\boldsymbol d}}\nc\bfD{{\boldsymbol D}}\nc\cD{{\mathcal D}}
\nc\bfe{{\boldsymbol e}}\nc\bfE{{\boldsymbol E}}\nc\cE{{\mathcal E}}
\nc\bff{{\boldsymbol f}}\nc\bfF{{\boldsymbol F}}\nc\cF{{\mathcal F}}\nc\sF{{\mathscr F}}
\nc\bfg{{\boldsymbol g}}\nc\bfG{{\boldsymbol G}}\nc\cG{{\mathcal G}}
\nc\bfh{{\boldsymbol h}}\nc\bfH{{\boldsymbol H}}\nc\cH{{\mathcal H}}
\nc\bfi{{\boldsymbol i}}\nc\bfI{{\boldsymbol I}}\nc\cI{{\mathcal I}}\nc\sI{{\mathscr I}}
\nc\bfj{{\boldsymbol j}}\nc\bfJ{{\boldsymbol J}}\nc\cJ{{\mathcal J}}
\nc\bfk{{\boldsymbol k}}\nc\bfK{{\boldsymbol K}}\nc\cK{{\mathcal K}}
\nc\bfl{{\boldsymbol l}}\nc\bfL{{\boldsymbol L}}\nc\cL{{\mathcal L}}
\nc\bfm{{\boldsymbol m}}\nc\bfM{{\boldsymbol M}}\nc\cM{{\mathcal M}}
\nc\bfn{{\boldsymbol n}}\nc\bfN{{\boldsymbol N}}\nc\cN{{\mathcal N}}
\nc\bfo{{\boldsymbol o}}\nc\bfO{{\boldsymbol O}}\nc\cO{{\mathcal O}}
\nc\bfp{{\boldsymbol p}}\nc\bfP{{\boldsymbol P}}\nc\cP{{\mathcal P}}\nc\eP{{\EuScriptP}}\nc\fP{{\mathfrak P}}
\nc\bfq{{\boldsymbol q}}\nc\bfQ{{\boldsymbol Q}}\nc\cQ{{\mathcal Q}}
\nc\bfr{{\boldsymbol r}}\nc\bfR{{\boldsymbol R}}\nc\cR{{\mathcal R}}\nc\sR{{\mathscr R}}
\nc\bfs{{\boldsymbol s}}\nc\bfS{{\boldsymbol S}}\nc\cS{{\mathcal S}}
\nc\bft{{\boldsymbol t}}\nc\bfT{{\boldsymbol T}}\nc\cT{{\mathcal T}}
\nc\bfu{{\boldsymbol u}}\nc\bfU{{\boldsymbol U}}\nc\cU{{\mathcal U}}
\nc\bfv{{\boldsymbol v}}\nc\bfV{{\boldsymbol V}}\nc\cV{{\mathcal V}}\nc\sV{{\mathscr V}}
\nc\bfw{{\boldsymbol w}}\nc\bfW{{\boldsymbol W}}\nc\cW{{\mathcal W}}\nc\sW{{\mathscr W}}
\nc\bfx{{\boldsymbol x}}\nc\bfX{{\boldsymbol X}}\nc\cX{{\mathcal X}}
\nc\bfy{{\boldsymbol y}}\nc\bfY{{\boldsymbol Y}}\nc\cY{{\mathcal Y}}
\nc\bfz{{\boldsymbol z}}\nc\bfZ{{\boldsymbol Z}}\nc\cZ{{\mathcal Z}}

\begin{document}

%\pagenumbering{gobble}

\title{Explicit constructions of high-rate MDS array codes with optimal repair bandwidth}

\author{\IEEEauthorblockN{Min Ye} \hspace*{2in}
\and \IEEEauthorblockN{Alexander Barg}}

\maketitle
{\renewcommand{\thefootnote}{}\footnotetext{

\vspace{-.2in}
 
\noindent\rule{1.5in}{.4pt}

The authors are with Dept. of ECE and ISR, University of Maryland, College Park, MD 20742. Emails: yeemmi@gmail.com and abarg@umd.edu.
Research supported by NSF grants CCF1422955 and CCF1217245.}
}
\renewcommand{\thefootnote}{\arabic{footnote}}
\setcounter{footnote}{0}

\begin{abstract}
Maximum distance separable (MDS) codes are optimal error-correcting codes in the sense that they provide the maximum failure-tolerance 
for a given number of parity nodes. Suppose that an MDS code with $k$ information nodes and $r=n-k$ parity nodes
is used to encode data in a distributed storage system. It is known that if $h$ out of the $n$ nodes are inaccessible and $d$ 
surviving (helper) nodes are used to recover the lost data, then we need to download at least $h/(d+h-k)$ fraction of the data stored in each of the helper nodes (Dimakis et al., 2010 and Cadambe et al., 2013). If this lower bound is achieved for the repair
of any $h$ erased nodes from any $d$ helper nodes, we say that the MDS code 
has the $(h,d)$-optimal repair property.

We study high-rate MDS array codes with the optimal repair property. Explicit constructions of such codes in the literature are only available for the cases where there are at most 3 parity nodes, and these existing constructions can only optimally repair a single node failure by accessing all the surviving nodes.

In this paper, given any $r$ and $n$, we present two explicit constructions of MDS array codes with the $(h,d)$-optimal repair property for all $h\le r$ and $k\le d\le n-h$ simultaneously. Codes in the first family
can be constructed over any base field $F$ as long as $|F|\ge sn,$ where $s=\text{lcm}(1,2,\dots,r).$ The 
encoding, decoding, repair of failed nodes, and update procedures of these codes all have low complexity.
Codes in the second family have the optimal access property and can be constructed over any base field $F$ as long as $|F|\ge n+1.$ Moreover, both code families have the optimal error resilience capability when repairing failed nodes. We also construct several other related families of MDS codes with the optimal repair property.

\end{abstract}

\section{Introduction}

Distributed storage systems, such as those run by Google \cite{Google03} and Facebook \cite{Facebook07}, are widely used for data storage, with applications ranging from social networks to file and video sharing. Currently deployed systems are formed
of thousands of individual drives (nodes), and drive failures occur on a daily basis.
For this reason, companies utilizing or providing distributed storage solutions have increasingly turned 
to error-correcting coding for efficient recovery of data stored in the system.
The coding method of choice used for data protection relies on MDS codes which provide the maximum failure tolerance for a given 
amount of storage overhead. The distributed nature of the system introduces new challenges in the
code design that are related to the need to communicate data between the nodes during the repair of node failures.
Efficient operation of the system requires minimizing the {\em repair bandwidth}, i.e., the amount of data 
that needs to be downloaded to repair the contents of the failed node(s). Therefore, recent research 
on MDS codes for distributed storage has focused on codes with minimum repair 
bandwidth. 

In this paper, following the literature on 
codes for storage, we use the terminology motivated by storage applications: code's coordinates are called nodes,
correcting erasures is referred to as repairing (or recovering) failed nodes, and
the process of obtaining the information stored at some other nodes in 
the codeword to repair the failed node(s) is described as ``downloading data from the helper nodes". 
As usual in distributed storage applications, we assume that an MDS code of length $n$ formed of $k$ information coordinates and 
$r=n-k$ parity coordinates is spread across $n$ different nodes of the storage cluster. Each node of the cluster stores a coordinate of the code.
%Suppose that $h\ge 1$ nodes become unavailable, and the system attempts to repair their
%contents by accessing $d\le n-h$ surviving (helper) nodes. In this case, as shown in 
%\cite{Dimakis10,Cadambe13}, the recovery requires to download
%at least an $h/(d+h-k)$ fraction of the data stored in each of the helper nodes.

\subsection{Notation and earlier results}
Most studies of MDS codes with optimal repair bandwidth in the literature are concerned with a particular subclass of codes known as MDS {\em array codes} \cite{Blaum98}. An $(n,k,l)$ 
MDS array code has $k$ information nodes and $r=n-k$ parity nodes in each codeword with the property that 
any $k$ out of $n$ nodes can recover the codeword.  Every
node is a column vector in $F^l,$ where $F$ is some finite field, reflecting the fact that the system
views a large data block stored in one node as one coordinate of the codeword.

Suppose that $h\ge 1$ nodes become unavailable, and the system attempts to repair their
contents by accessing $d\le n-h$ surviving (helper) nodes. 
In this case, as shown in 
\cite{Dimakis10,Cadambe13}, the recovery of failed nodes requires to download
at least an $h/(d+h-k)$ fraction of the data stored in each of the helper nodes. More formally, 
given an $(n,k,l)$ MDS array code $\cC$ and two disjoint subsets ${\sF},{\sR}\subseteq[n]$ such that $|{\sF}|\le r$ and $|{\sR}|\ge k,
$ we define $N(\cC,{\sF},{\sR})$ as the smallest number of symbols\footnote{these symbols can be some functions of the contents of the 
nodes $\{C_i,i\in{\sR}\}.$} of $F$ one needs to download from the surviving nodes $\{C_i,i\in{\sR}\}$ in order to recover the erased 
nodes $\{C_i,i\in{\sF}\}.$ (We use the notation $[n]=\{1,2,\dots,n\}.$) Dimakis et. al. \cite{Dimakis10} showed that 
\begin{equation}\label{eq:lowerb}
N(\cC,{\sF},{\sR})\ge \frac{|{\sF}||{\sR}|l}{|{\sF}|+|{\sR}|-k}
\end{equation}
for any $(n,k,l)$ MDS array code $\cC$ and any two disjoint subsets ${\sF},{\sR}\subseteq[n]$ such that $|{\sF}|=1$ and $|{\sR}|\ge k.$ Cadambe et. al. \cite{Cadambe13} further proved that \eqref{eq:lowerb} holds for any two disjoint subsets ${\sF},{\sR}\subseteq[n]$ such that $|{\sF}|\le r$ and $|{\sR}|\ge k.$ If \eqref{eq:lowerb} is achieved, we say that $\cC$ can \emph{optimally repair} nodes $\{C_i,i\in{\sF}\}$ using nodes $\{C_i,i\in{\sR}\}.$

 Given an $(n,k,l)$ MDS array code $\cC,$ we define the {\em $(h,d)$-repair bandwidth} of $\cC$ as
$$
\max_{{\sF}\bigcap{\sR}=\emptyset, |{\sF}|=h,|{\sR}|=d} N(\cC,{\sF},{\sR}).
$$
If the $(h,d)$-repair bandwidth of $\cC$ is $\frac{hdl}{h+d-k},$ meeting the lower bound in \eqref{eq:lowerb}, we say that $\cC$ has {\em $(h,d)$-optimal repair} property. We further say that the code has the {\em $d$-optimal repair} property 
if $h=1$, omitting the reference to $h$, and say that the code has the \emph{optimal repair} property if $h=1$ and $d=n-1.$

In \cite{Pawar11,Rashmi12}, the authors considered the \emph{error resilience} capability in the repair process of a single node failure. We generalize this concept to the repair process of multiple node failures. 
Given an $(n,k,l)$ MDS array code $\cC,$ a nonnegative integer $t$ and two disjoint subsets ${\sF},{\sR}\subseteq[n]$ such that $|{\sF}|\le r$ and $|{\sR}|\ge k+2t,$ we define $N(\cC,{\sF},{\sR},t)$ as the smallest number of symbols of $F$ one needs to download from nodes $\{C_i,i\in{\sR}\}$ such that the erased nodes $\{C_i,i\in{\sF}\}$ can be recovered from these symbols as long as the number of erroneous nodes in $\{C_i,i\in{\sR}\}$ is no larger than $t.$ It is shown in \cite{Pawar11,Rashmi12} that
    \begin{equation}\label{eq:lberr}
N(\cC,{\sF},{\sR},t)\ge \frac{|{\sF}||{\sR}|l}{|{\sF}|+|{\sR}|-2t-k}
    \end{equation}
for any $(n,k,l)$ MDS array code $\cC,$ any nonnegative integer $t$ and any two disjoint subsets ${\sF},{\sR}\subseteq[n]$ such that $|{\sF}|=1$ and $|{\sR}|\ge k+2t.$ The method in \cite{Pawar11,Rashmi12} can be straightforwardly generalized to show that \eqref{eq:lberr} also holds for any ${\sF}$ such that $|{\sF}|\le r.$ We say that an $(n,k,l)$ MDS array code $\cC$ has the {\em universally error-resilient (UER) $(h,d)$-optimal repair} property if
$$
N(\cC,{\sF},{\sR},t)= \frac{h(d+2t)l}{h+d-k}
$$
for any nonnegative integer $t$ and any two disjoint subsets ${\sF},{\sR}\subseteq[n]$ such that $|{\sF}|=h$ and $|{\sR}|= d+2t.$ As 
above, when $h=1,$ we omit it from the notation.

In general, the downloaded data in the repair process can be some functions, e.g. , linear combinations of the data stored in the helper nodes. As a 
result, even for codes with optimal repair property, we might still need to access a larger amount of data than the lower bounds in 
\eqref{eq:lowerb} and \eqref{eq:lberr}. We say that an $(n,k,l)$ MDS array code $\cC$ has {\em $(h,d)$-optimal access} property if 
the repair of any $h$ erased nodes using any $d$ helper nodes can be accomplished by accessing the amount of data that meets the lower bound in \eqref{eq:lowerb}.
We define the {\em UER  $(h,d)$-optimal access} property in a similar way.

For $k\le (n+1)/2$ (the low rate regime), MDS array codes with $d$-optimal repair property were constructed in 
\cite{Rashmi11,Rashmi14,Shah12,Suh11,Wu09}, and MDS array codes with the UER $d$-optimal repair property were constructed in 
\cite{Rashmi12}. For arbitrary code rate, \cite{Cadambe13} proved that there exists a family of codes for which the bound 
\eqref{eq:lowerb} is asymptotically achieved when $l\to\infty$. For finite $l$ and $k>(n+1)/2$ (the high-rate regime)
papers \cite{Cadambe11,Cadambe11a,Papai13,Tamo13,Wang14} showed that for large enough base field $F$ there exist MDS array codes that can 
optimally repair any single systematic node failure using all the surviving nodes, and \cite{Wang11} showed the same for all rather than 
only systematic nodes.

{Among the very recent additions to the literature,
\cite{Rawat16} showed existence of MDS array codes with $d$-optimal repair property for any single 
value of $d$ in the range $k\le d\le n-1.$
Paper \cite{Goparaju16} showed existence of MDS array codes that can optimally repair any single systematic node 
failure by accessing any subset of the surviving nodes as long as the number of accessed nodes is no less than $k.$
Finally \cite{Wang16} showed existence of MDS array codes which can optimally repair any $h$ systematic node failures using all 
the $n-h$ surviving nodes for any single value of $h$ in the range $1\le h\le r$ (some special cases for $h=2$ and $3$ are discussed in \cite{Rawat16a}). These papers essentially rely on existential lemmas
in large finite fields, e.g., the Schwartz-Zippel lemma or Combinatorial Nullstellensatz. 
At the same time, all the known explicit constructions in the high-rate regime are obtained only for the case of at most $3$ parity nodes, and are further limited to repairing $h=1$ failed node by accessing all the $n-1$ surviving nodes; see
\cite{Papai13,Wang14,Tamo13,Raviv15,Wang11}.}

\subsection{Overview of the paper}
In this paper, given any $n$ and $k,$ we present two explicit families of $(n,k,l)$ MDS array codes.
Codes in the first family have the UER $(h,d)$-optimal repair property for all $h\le r$ and $k\le d\le n-h$ simultaneously, and the encoding, decoding, repair and update procedures of these codes all have low complexity.
Codes in the second family have the UER $(h,d)$-optimal access property for all $h\le r$ and $k\le d\le n-h$ simultaneously, and can be constructed over any base field $F$ as long as $|F|\ge n+1.$

The paper is organized as follows. In Sections \ref{opt}-\ref{1ult} we present the first code family and its variants.
Sections \ref{permut}-\ref{2ult} are devoted to the second family and related results. Finally, in Sect.~\ref{GRSA}
we present a new class of MDS array codes with optimal repair and smaller $l$ than the other constructions. In the next
few paragraphs we give a more detailed overview of our results.

In Section \ref{opt}, we present our first construction of $(n,k,l=r^n)$ MDS array codes
with the optimal repair property using any field $F$ of size $|F|\ge rn$. The encoding, decoding, and repair of a single failed node involve only simple operations with $r\times r$ matrices over $F$, and thus have low complexity. An additional property of the proposed codes is optimal update, i.e., the need to change only the minimum possible number of coordinates in the parity nodes if 
one coordinate in systematic node is updated. In our construction we rely on a (non-systematic) parity-check representation of the 
codes as opposed to the systematic generator form used in most earlier works. This representation does not 
distinguish between systematic nodes and parity nodes, and leads naturally to the optimal repair of all nodes. 
The parity-check form combined with the block Vandermonde structure \cite{Wang14} and the idea of using $r$-ary expansions 
\cite{Cadambe11,Tamo13} makes the explicit construction for larger number of parity nodes possible. Note that exponentially large $l$ 
is necessary for optimal repair bandwidth: Indeed, according to a result of \cite{Tamo14a}, $l\ge 2^{\sqrt {k/(2r-1)}}$ is necessary
for any code with the optimal repair property. It is further shown in \cite{Tamo14} that $l\ge r^{(k-1)/r}$ for any code with the optimal access property.

In Section \ref{dopt}, we extend the construction of Section \ref{opt} to obtain $(n,k,l)$ MDS array codes with the UER $d$-optimal repair property for any positive integers $n,k,d,l$ such that $k\le d<n,l=(d+1-k)^n$ using any field $F$ of size $|F|\ge (d+1-k)n$. We first observe that we only need to know  a $1/(d+1-k)$ fraction of data stored in each of the surviving nodes $C_j,j\neq i$ in order to recover the erased node $C_i.$ We then use a novel method to prove that these data form an 
$(n-1,d,l/(d+1-k))$ MDS array code, and thus establish the UER $d$-optimal repair property. 

In Section \ref{simul}, we present another extension, constructing MDS codes with $d$-optimal repair property for several 
values of $d$ simultaneously. Moreover, we show that $(n,k,r^n)$ MDS array codes constructed in Section \ref{opt} will automatically have the $d$-optimal repair property for all $d$ such that $(d+1-k)|(n-k).$
In Section \ref{1ult} we further extend our construction to obtain $(n,k,l)$ MDS array codes with the UER 
$(h,d)$-optimal repair property for all $h\le r$ and $k\le d\le n-h$ simultaneously, where $l=s^n,s=\text{lcm}(1,2,\dots,r).$ 
These codes also have the optimal update property. Moreover, the encoding, decoding, and repair procedures only require operations with matrices of size not greater than $n\times n.$

In Section \ref{permut} we develop the idea of using permutation matrices \cite{Cadambe11,Tamo13} 
 to obtain an explicit family of $(n,n-r,r^{n-1})$ MDS array codes with the optimal access property, which
can be constructed over any base field $F$ such that $|F|\ge n+1.$ In Sections \ref{2d}-\ref{2ult}, we combine the ideas in Section \ref{permut} and in Sections \ref{dopt}-\ref{1ult} to obtain an explicit family of $(n,k,l)$ MDS array codes with the UER $(h,d)$-optimal access property for all $h\le r$ and $k\le d\le n-h$ simultaneously, where $l=s^n,s=\text{lcm}(1,2,\dots,r).$ These codes can be constructed over any base field $F$ as long as $|F|\ge n+1.$

In Section \ref{GRSA} we introduce a new class of MDS array codes, {\em Generalized Reed-Solomon Array Codes}, and 
use their properties to extend the construction of Section \ref{permut} to obtain MDS array codes with the 
$d$-optimal repair property for several values of $d$ simultaneously. These codes also only require the underlying 
field size $|F|\ge n+1$ as well as a smaller $l$ compared to the other code families in this paper.
%constructions in Sections \ref{dopt}-\ref{simul} and Sections \ref{2d}-\ref{2md}.

\section{General code construction} Let $\cC\in F^{ln}$ be an $(n,k,l)$ array code with nodes $C_i\in F^l, i=1,\dots,n,$ where each $C_i$ is a column vector.
Throughout this paper we consider codes defined in the following parity-check form:
\begin{equation}\label{eq:parityform}
\cC=\{(C_1,C_2,\dots,C_n):\sum_{i=1}^n A_{t,i}C_i=0,\, t=1,\dots,r\},
\end{equation}
where $A_{t,i}, t=1,\dots,r, i=1,\dots, n$ are $l\times l$ matrices over $F$.  

Given positive integers $r$ and $n,$ define an $(n,k=n-r,l)$ array code $\cC$ by setting in \eqref{eq:parityform}
\begin{equation}\label{eq:power}
A_{t,i}=A_i^{t-1}, t\in[r],i\in[n],
\end{equation}
where $A_1,A_2,\dots,A_n$ are some $l\times l$ matrices. (We use the convention $A^0=I.$)
The specific code families in Section \ref{opt}-\ref{GRSA} are obtained by choosing different forms of the matrices $A_1,A_2,\dots,A_n.$

\section{Construction of MDS array codes with optimal repair property}\label{opt}
%We use the notation $[n]=\{1,2,\dots,n\}.$ 

\subsection{Code construction}

\begin{cnstr}\label{constr1}   Let $F$ be a finite field of size $|F|\ge rn,$ and let $l=r^n.$ Let  $\{\lambda_{i,j}\}_{i\in[n],j=0,1,\dots,r-1}$ be $rn$ distinct elements in $F.$  
Consider the code family given by \eqref{eq:parityform}-\eqref{eq:power}, where we take
$$
  A_i=\sum_{a=0}^{l-1}\lambda_{i,a_i}e_a e_a^T
   ,\; i=1,\dots,n.
$$
Here $\{e_a:a=0,1,\dots,l-1\}$ is the standard basis of $F^l$ over $F,$ and $a_i$ is the $i$-th digit from the
right in the representation of $a$ in the $r$-ary form, $a=(a_n,a_{n-1},
\dots,a_1).$ 
\end{cnstr}

Since the $A_i,i=1,\dots,n$ are diagonal matrices, we can write out the parity-check equations \eqref{eq:parityform} coordinatewise. 
Let $c_{i,a}$ denote the $a$-th coordinate of the column vector $C_i$ for all $a=0,\dots,l-1,$ i.e.,
$C_i=(c_{i,0},c_{i,1}, \dots, c_{i,l-1})^T.$
We have
\begin{equation}\label{eq:scalar}
\sum_{i=1}^n \lambda_{i,a_i}^t c_{i,a}=0
\end{equation}
for all $t=0,\dots,r-1$ and $a=0,\dots,l-1.$

\begin{theorem} Codes given by
Construction \ref{constr1} attain optimal repair bandwidth for repairing any single failed node.
\end{theorem}
\begin{IEEEproof} For $u=0,1,\dots,r-1,$ let 
$a(i,u):=(a_n,\dots,a_{i+1},u,a_{i-1},\dots,a_1).$
We will show that for any $i\in[n]$ and $a=0,1,\dots,l-1,$ the coordinates $\{c_{i,a(i,0)},c_{i,a(i,1)},\dots,c_{i,a(i,r-1)}\}$ in $C_i$
are functions of the following set of $n-1$ elements of $F$:
  \begin{equation}\label{eq:mu}
   \mu^{(a)}_{j,i}:=\sum_{u=0}^{r-1}c_{j,a(i,u)}, \quad j\in [n]\backslash\{i\}. 
  \end{equation}
In other words, each surviving node only needs to transmit one scalar in $F$ to recover $r$ coordinates in the failed node, so the optimal repair bandwidth is achieved.
%we can calculate $r$ coordinates of $C_i$  by downloading only a scalar $\sum_{u=0}^{r-1}c_{j,a(i,u)}$ in $F$ from each surviving node $C_j,j\neq i.$ Thus we can achieve the optimal repair bandwidth when repairing any single node failure. 
Replacing $a$ with $a(i,u)$ in \eqref{eq:scalar}, we obtain
\begin{equation}\label{eq:jneqi}
\lambda_{i,u}^tc_{i,a(i,u)}+\sum_{j\neq i}\lambda_{j,a_j}^tc_{j,a(i,u)}=0.
\end{equation}
Summing \eqref{eq:jneqi} over $u=0,1,\dots,r-1$ and then writing the result in matrix form, we get
\begin{equation}\label{eq:singlerepair}
\begin{aligned}
&\left[\!\! \begin{array}{cccc} 1 & 1 & \dots & 1\\
\lambda_{i,0} & \lambda_{i,1} & \dots & \lambda_{i,r-1}\\
\vdots & \vdots & \vdots & \vdots\\
\lambda_{i,0}^{r-1} & \lambda_{i,1}^{r-1} & \dots & \lambda_{i,r-1}^{r-1} \end{array} \!\!\right]
\left[ \!\!\begin{array}{c} c_{i,a(i,0)}\\
c_{i,a(i,1)}\\
\vdots \\
c_{i,a(i,r-1)}\\ \end{array} \!\!\right] 
%\\&\hspace*{.2in}
=-\left[ \begin{array}{c} \sum_{j\neq i}\mu^{(a)}_{j,i}\\
\sum_{j\neq i}\lambda_{j,a_j}\mu^{(a)}_{j,i}\\
\vdots \\
\sum_{j\neq i}\lambda_{j,a_j}^{r-1}\mu^{(a)}_{j,i} \end{array} \right] .
\end{aligned}
\end{equation}
By construction $\lambda_{i,0},\dots,\lambda_{i,r-1}$ are distinct, so we can solve this system for $\{c_{i,a(i,0)},c_{i,a(i,1)},\dots,c_{i,a(i,r-1)}\}$ given the set of elements in \eqref{eq:mu}. %values of $\{\mu^{(a)}_{ji}\}_{j\neq i}.$
\end{IEEEproof}
The repair procedure of a single node has low complexity: indeed, according to \eqref{eq:singlerepair}, it can be 
accomplished by operations with $r\times r$ matrices (rather than much larger $l\times l$ matrices).

\begin{theorem}\label{1stmds}
The code $\cC$ given by Construction \ref{constr1} is MDS.
\end{theorem}
\begin{IEEEproof}
We write out the parity-check equations \eqref{eq:parityform} coordinatewise. For all $a=0,1,\dots,l-1,$ we have
\begin{equation}\label{eq:d-opt}
\left[ \begin{array}{cccc} 1 & 1 & \dots & 1\\
\lambda_{1,a_1} & \lambda_{2,a_2} & \dots & \lambda_{n,a_n}\\
\vdots & \vdots & \vdots & \vdots\\
\lambda_{1,a_1}^{r-1} & \lambda_{2,a_2}^{r-1} & \dots & \lambda_{n,a_n}^{r-1} \end{array} \right]
\left[ \begin{array}{c} c_{1,a}\\
c_{2,a}\\
\vdots \\
c_{n,a}\\ \end{array} \right]=0
\end{equation}
Clearly every $r$ columns of the parity-check matrix in \eqref{eq:d-opt} have rank $r$, so any $k$ out of $n$ elements in the set 
$\{c_{1,a},c_{2,a},\dots,c_{n,a}\}$ can recover the whole set. Since this holds for all $a=0,1,\dots,l-1,$ we conclude that any $k$ nodes of a codeword in $\cC$ can recover the whole codeword. 
\end{IEEEproof}

\subsection{Complexity of encoding, decoding, and updates}
The code given by Construction \ref{constr1} can be efficiently transformed into systematic form. Without loss of generality we assume that the first $k$ nodes are systematic (information) nodes. By \eqref{eq:d-opt}, for all $a=0,1,\dots,l-1,$ we have
\begin{equation}\label{eq:systematic}
\begin{aligned}
&\left[ \begin{array}{cccc} 1 & 1 & \dots & 1\\
\lambda_{k+1,a_{k+1}} & \lambda_{k+2,a_{k+2}} & \dots & \lambda_{k+r,a_{k+r}}\\
\vdots & \vdots & \vdots & \vdots\\
\lambda_{k+1,a_{k+1}}^{r-1} & \lambda_{k+2,a_{k+2}}^{r-1} & \dots & \lambda_{k+r,a_{k+r}}^{r-1} \end{array} \right]
\left[ \begin{array}{c} c_{k+1,a}\\
c_{k+2,a}\\
\vdots \\
c_{k+r,a}\\ \end{array} \right] =-\left[ \begin{array}{cccc} 1 & 1 & \dots & 1\\
\lambda_{1,a_1} & \lambda_{2,a_2} & \dots & \lambda_{k,a_k}\\
\vdots & \vdots & \vdots & \vdots\\
\lambda_{1,a_1}^{r-1} & \lambda_{2,a_2}^{r-1} & \dots & \lambda_{k,a_k}^{r-1} \end{array} \right]
\left[ \begin{array}{c} c_{1,a}\\
c_{2,a}\\
\vdots \\
c_{k,a}\\ \end{array} \right] .
\end{aligned}
\end{equation}
Consequently, in the encoding process we do not need to invert an $rl\times rl$ matrix, instead, we only need to invert $r\times r$ matrices $l$ times, gaining a factor of $l^2$ in complexity. Similarly, in the decoding process, if some $r$ nodes are erased, then in order
to recover them, we only need to invert $r\times r$ matrices $l$ times.

Another useful parameter of codes is {\em update complexity} \cite{Blaum98}. On account of the MDS property, in order to update the value of a stored element $c_{i,a}$ in an information node, one needs to update at least one coordinate in every parity node \cite{Tamo14}. 
From \eqref{eq:systematic} it is easy to see that for any $i\in[k]$ and $a=0,\dots,l-1,$ to update $c_{i,a},$ we only need to update $c_{k+1,a},\dots,c_{k+r,a}.$ Thus Construction \ref{constr1} gives an optimal update code.

\section{Explicit MDS array codes with the UER $d$-optimal repair property}\label{dopt}
The general construction in \eqref{eq:parityform}-\eqref{eq:power} can also be used to construct 
an $(n,k=n-r,l)$ MDS array code $\cC$ with the UER $d$-optimal repair property, $k\le d\le n-1.$

\begin{cnstr}\label{constr2}
Let $F$ be a finite field of size $|F|\ge sn,$ where $s=d+1-k.$
Let $\{\lambda_{i,j}\}_{i\in[n],j=0,1,\dots,s-1}$ be $sn$ distinct elements in $F.$
Consider the code family given by \eqref{eq:parityform}-\eqref{eq:power}, where $l=s^n$ and
$$
  A_i=\sum_{a=0}^{l-1}\lambda_{i,a_i}e_a e_a^T
   ,\; i=1,\dots,n.
$$
Here $\{e_a:a=0,1,\dots,l-1\}$ is the standard basis of $F^l$ over $F$ and $a_i$ is the $i$-th digit from the right
in the representation of $a$ in the $s$-ary form, $a=(a_n,a_{n-1},
\dots,a_1).$ 
\end{cnstr}

Define $a(i,u)$ and $c_{i,a}$ in the same way as in Sect.~\ref{opt}.

\begin{theorem}\label{oldmds} 
The code $\cC$ given by Construction \ref{constr2} is an MDS code.
\end{theorem}
\begin{IEEEproof}
Same as the proof of Theorem \ref{1stmds}.
\end{IEEEproof}

By the same arguments as in the previous section, $\cC$ also has low-complexity encoding, decoding, and the optimal update property.

Let us show that the code $\cC$ has the UER $d$-optimal repair property. Recall the definition of Generalized Reed-Solomon codes.
\begin{definition}
A \emph{Generalized Reed-Solomon code} $\text{\rm GRS}(n,k,\Omega,v)\subseteq F^n$ of dimension $k$ over $F$ 
with evaluation points $\Omega=\{\omega_1,\omega_2,\dots,\omega_n\}\subseteq F$  is the set of vectors
\begin{align*}
\{(v_1f(\omega_1),\dots,v_nf(\omega_n))\in F^n:f\in F[x], \deg f\le k-1\}
\end{align*}
where $v=(v_1,\dots,v_n)\in (F^\ast)^n$ are some nonzero coefficients. If $v=(1,\dots,1),$ then the GRS code is called
a Reed-Solomon code.
%
%A \emph{Reed-Solomon code} $\text{\rm RS}(n,k,\Omega)\subseteq F^n$ of dimension $k$ over a finite field $F$ with evaluation points $\Omega=\{\omega_1,\omega_2,\dots,\omega_n\}\subseteq F$ is the set of vectors
%\begin{align*}
%\{(f(\omega_1),\dots,f(\omega_n))\in F^n: f\in F[x], \deg f\le k-1\}.
%\end{align*}
\end{definition}
\begin{remark}\label{r:GRS}  The minimum distance of the code $\text{\rm GRS}(n,k,\Omega,v)$ is $n-k+1.$ 
Note that the projection of the GRS code on any subset of coordinates $\Omega'\subset \Omega, |\Omega'|\ge k,$ is itself a GRS code. In particular,
its distance equals $|\Omega'|-k+1.$
\end{remark}

\begin{theorem}\label{oldoptimal}
The code $\cC$ given by Construction \ref{constr2} has the UER $d$-optimal repair property.
\end{theorem}
\begin{IEEEproof}
Without loss of generality, we consider the case of repairing $C_1.$ Let
  \begin{equation}\label{eq:dmu}
   \mu^{(a)}_{j,1}:=\sum_{u=0}^{s-1}c_{j,a(1,u)}, \quad j\in \{2,3,\dots,n\}. 
  \end{equation}
Using arguments similar to those that lead to \eqref{eq:singlerepair}, we obtain
\begin{equation}\label{eq:dsr}
\begin{aligned}
&\left[\!\! \begin{array}{cccc} 1 & 1 & \dots & 1\\
\lambda_{1,0} & \lambda_{1,1} & \dots & \lambda_{1,s-1}\\
\lambda_{1,0}^2 & \lambda_{1,1}^2 & \dots & \lambda_{1,s-1}^2\\
\vdots & \vdots & \vdots & \vdots\\
\lambda_{1,0}^{r-1} & \lambda_{1,1}^{r-1} & \dots & \lambda_{1,s-1}^{r-1} \end{array} \!\!\right]
\left[ \!\!\begin{array}{c} c_{1,a(1,0)}\\
c_{1,a(1,1)}\\
\vdots \\
c_{1,a(1,s-1)}\\ \end{array} \!\!\right] 
%\\&
=-\left[\!\! \begin{array}{cccc} 1 & 1 & \dots & 1\\
\lambda_{2,a_2} & \lambda_{3,a_3} & \dots & \lambda_{n,a_n}\\
\lambda_{2,a_2}^2 & \lambda_{3,a_3}^2 & \dots & \lambda_{n,a_n}^2\\
\vdots & \vdots & \vdots & \vdots\\
\lambda_{2,a_2}^{r-1} & \lambda_{3,a_3}^{r-1} & \dots & \lambda_{n,a_n}^{r-1} \end{array} \!\!\right]\left[ \begin{array}{c} \mu^{(a)}_{2,1}\\
\mu^{(a)}_{3,1}\\
\vdots \\
\mu^{(a)}_{n,1} \end{array} \right] .
\end{aligned}
\end{equation}
Define polynomials $p_0(x)=\prod_{u=0}^{s-1}(x-\lambda_{1,u}),$ and $p_i(x)=x^ip_0(x)$ for $i=0,1,\dots,r-s-1.$ We have proved the case of $d=n-1$ in the previous section, so here we only consider the case when $d<n-1,$ and so $r-s-1\ge 0.$ Since the degree of $p_i(x)$ is less than $r$ for all $i=0,1,\dots,r-s-1,$ we can write
$$
p_i(x)=\sum_{j=0}^{r-1}p_{i,j}x^j.
$$
Define the $(r-s)\times r$ matrix
$$
P=\left[\!\! \begin{array}{cccc} p_{0,0} & p_{0,1} & \dots & p_{0,r-1}\\
p_{1,0} & p_{1,1} & \dots & p_{1,r-1}\\
\vdots & \vdots & \vdots & \vdots\\
p_{r-s-1,0} & p_{r-s-1,1} & \dots & p_{r-s-1,r-1} \end{array} \!\!\right].
$$
Since
\begin{align*}
P
\left[\!\! \begin{array}{cccc} 1 & 1 & \dots & 1\\
\lambda_{1,0} & \lambda_{1,1} & \dots & \lambda_{1,s-1}\\
\lambda_{1,0}^2 & \lambda_{1,1}^2 & \dots & \lambda_{1,s-1}^2\\
\vdots & \vdots & \vdots & \vdots\\
\lambda_{1,0}^{r-1} & \lambda_{1,1}^{r-1} & \dots & \lambda_{1,s-1}^{r-1} \end{array} \!\!\right]
%\\
&=
\left[\!\! \begin{array}{cccc} p_0(\lambda_{1,0}) & p_0(\lambda_{1,1}) & \dots & p_0(\lambda_{1,s-1})\\
p_1(\lambda_{1,0}) & p_1(\lambda_{1,1}) & \dots & p_1(\lambda_{1,s-1})\\
\vdots & \vdots & \vdots & \vdots\\
p_{r-s-1}(\lambda_{1,0}) & p_{r-s-1}(\lambda_{1,1}) & \dots & p_{r-s-1}(\lambda_{1,s-1}) \end{array} \!\!\right]
=0,
\end{align*}
together with \eqref{eq:dsr}, we have
\begin{equation}\label{eq:pol}
P\left[\!\! \begin{array}{cccc} 1 & 1 & \dots & 1\\
\lambda_{2,a_2} & \lambda_{3,a_3} & \dots & \lambda_{n,a_n}\\
\lambda_{2,a_2}^2 & \lambda_{3,a_3}^2 & \dots & \lambda_{n,a_n}^2\\
\vdots & \vdots & \vdots & \vdots\\
\lambda_{2,a_2}^{r-1} & \lambda_{3,a_3}^{r-1} & \dots & \lambda_{n,a_n}^{r-1} \end{array} \!\!\right]\left[ \begin{array}{c} \mu^{(a)}_{2,1}\\
\mu^{(a)}_{3,1}\\
\vdots \\
\mu^{(a)}_{n,1} \end{array} \right] =0.
\end{equation}

Note that 
\begin{align*}
P\left[\!\! \begin{array}{cccc} 1 & 1 & \dots & 1\\
\lambda_{2,a_2} & \lambda_{3,a_3} & \dots & \lambda_{n,a_n}\\
\lambda_{2,a_2}^2 & \lambda_{3,a_3}^2 & \dots & \lambda_{n,a_n}^2\\
\vdots & \vdots & \vdots & \vdots\\
\lambda_{2,a_2}^{r-1} & \lambda_{3,a_3}^{r-1} & \dots & \lambda_{n,a_n}^{r-1} \end{array} \!\!\right]
&=\left[\!\! \begin{array}{cccc} p_0(\lambda_{2,a_2}) & p_0(\lambda_{3,a_3}) & \dots & p_0(\lambda_{n,a_n})\\
p_1(\lambda_{2,a_2}) & p_1(\lambda_{3,a_3}) & \dots & p_1(\lambda_{n,a_n})\\
\vdots & \vdots & \vdots & \vdots\\
p_{r-s-1}(\lambda_{2,a_2}) & p_{r-s-1}(\lambda_{3,a_3}) & \dots & p_{r-s-1}(\lambda_{n,a_n}) \end{array} \!\!\right]\\
&\hspace*{-1in}=\left[\!\! \begin{array}{cccc} p_0(\lambda_{2,a_2}) & p_0(\lambda_{3,a_3}) & \dots & p_0(\lambda_{n,a_n})\\
p_0(\lambda_{2,a_2})\lambda_{2,a_2} & p_0(\lambda_{3,a_3})\lambda_{3,a_3} & \dots & p_0(\lambda_{n,a_n})\lambda_{n,a_n}\\
\vdots & \vdots & \vdots & \vdots\\
p_0(\lambda_{2,a_2})\lambda_{2,a_2}^{r-s-1} & p_0(\lambda_{3,a_3})\lambda_{3,a_3}^{r-s-1} & \dots & p_0(\lambda_{n,a_n})\lambda_{n,a_n}^{r-s-1} \end{array} \!\!\right].
\end{align*}
Moreover, $p_0(\lambda_{2,a_2}), p_0(\lambda_{3,a_3}), \dots , 
p_0(\lambda_{n,a_n})$ are all nonzero. Thus $(\mu_{2,1}^{(a)},\mu_{3,1}^{(a)},\dots,\mu_{n,1}^{(a)})$ forms a Generalized 
Reed-Solomon code of length $n-1$ and dimension $d.$ 
According to Remark \ref{r:GRS}, given a nonnegative integer $t$ such that $d+2t<n,$ any $d+2t$ out of $n-1$ 
elements in $\{\mu_{2,1}^{(a)},\mu_{3,1}^{(a)},\dots,\mu_{n,1}^{(a)}\}$ suffice to recover the whole set as long as the number of 
erroneous elements in the $d+2t$ elements is not greater than $t.$ 
Moreover, \eqref{eq:dsr} implies that  $\{c_{1,a(1,0)},c_{1,a(1,1)},
\dots,c_{1,a(1,s-1)}\}$ can be determined by $\{\mu_{2,1}^{(a)},\mu_{3,1}^{(a)},\dots,\mu_{n,1}^{(a)}\}.$ 
Consequently, we can recover 
$C_1$ by accessing any $d+2t$ surviving nodes and downloading the total of $(d+2t)l/s$ symbols of $F$ from these nodes as long as the 
number of erroneous nodes among the helper nodes is not greater than $t.$ This completes the proof.
\end{IEEEproof}

\section{MDS array codes with the UER $d$-optimal repair property for several values of $d$ simultaneously}\label{simul}

In the previous two sections, we constructed MDS array codes with the UER $d$-optimal repair property for a single value of $d.$ In this section we give a simple extension of the previous constructions to make the code have the UER $d$-optimal repair property for several values of $d$ simultaneously. Let $n,k,m,d_1,d_2,\dots,d_m$ be any positive integers such that $k\le d_1,\dots,d_m<n.$ 
We will show that by replacing $s$ in Construction \ref{constr2} with the value
$$
s=\text{lcm}(d_1+1-k,d_2+1-k,\dots,d_m+1-k)
$$
we obtain an $(n,k,l=s^n)$  MDS array code $\cC$ with the UER $d_i$-optimal repair property for all $i=1,\dots,m$ simultaneously.

By Theorem \ref{oldmds}, $\cC$ is an MDS array code. In the next theorem we establish results about 
the repair properties of the code $\cC.$
%Let us prove that $\cC$ has the UER $d_i$-optimal repair property for any $i\in[m].$

\begin{theorem}\label{uer}
The code $\cC$ has the UER $d_i$-optimal repair property for any $i\in[m].$
\end{theorem}
\begin{IEEEproof}
%Given any $i\in[m],$ we show that $\cC$ has the UER $d_i$-optimal repair property. 
Let $s_i=d_i+1-k.$ Similarly to the proof of Theorem \ref{oldoptimal}, we only prove the case of repairing $C_1.$  Since $s_i|s,$ we 
can partition the set $\{0,1,\dots,s-1\}$ into $s/s_i$ subsets $\cI_1,\cI_2,\dots,\cI_{s/s_i},$ such that $|\cI_j|=s_i$ for any $j
\in[s/s_i],$ where $[s/s_i]=\{1,2,\dots,s/s_i\}.$ Following the proof of Theorem \ref{oldoptimal}, we can show that for 
any $j\in[s/s_i], a\in
\{0,1,\dots,l-1\},$ any nonnegative integer $t$ such that $d_i+2t<n,$ and any $\sR\subseteq\{2,3,\dots,n\}$ of size $|
\sR|=d_i+2t,$ we can recover 
$\{c_{1,a(1,u)}:u\in\cI_j\}$ by acquiring the set of values $\{\sum_{u\in\cI_j}c_{v,a(1,u)}:v\in\sR\}$ as long as the number of erroneous nodes in $\{C_i:i\in\sR\}$ is not greater than $t.$ 
Therefore, we can recover $C_1$ by accessing any $d_i+2t$ surviving nodes and downloading the total of $(d_i+2t)l/s_i$ symbols of $F$ from these nodes as long as the number of erroneous nodes in the helper nodes is not greater than $t.$ This completes the proof.
\end{IEEEproof}

\begin{corollary}
The $(n,k,(n-k)^n)$ MDS array code given by Construction \ref{constr1} has the UER $d$-optimal repair property if $(d+1-k)|(n-k).$
\end{corollary}

\begin{example}
A $(k+4,k,4^{k+4})$ MDS array code given by Construction \ref{constr1} will automatically have the UER $(k+1)$-optimal 
repair property. A $(k+6,k,6^{k+6})$ MDS array code given by Construction \ref{constr1} has both the UER $(k+1)$-optimal 
repair property and the UER $(k+2)$-optimal repair property.
\end{example}

\section{Explicit MDS array codes with the UER $(h,d)$-optimal repair property for all $h\le r$ and $k\le d\le n-h$ simultaneously}\label{1ult}

Given integers $n$ and $r,$
we construct a family of $(n,k=n-r,l)$ MDS array codes 
with the UER $(h,d)$-optimal repair property for all $h\le r$ and $k\le d\le n-h$ simultaneously.

\begin{cnstr}\label{constr3}
 Let $F$ be a finite field of size $|F|\ge sn,$ where $s=\text{lcm}(1,2,\dots,r).$ 
Let $\{\lambda_{i,j}\}_{i\in[n],j=0,1,\dots,s-1}$ be $sn$ distinct elements in $F.$  Let $l=s^n.$
Consider the code family given by \eqref{eq:parityform}-\eqref{eq:power}, where the matrices $A_i$ are 
given by
  \begin{equation}\label{eq:Ai3}
  A_i=\sum_{a=0}^{l-1}\lambda_{i,a_i}e_a e_a^T
   ,\; i=1,\dots,n.
  \end{equation}
Here $\{e_a:a=0,1,\dots,l-1\}$ is the standard basis of $F^l$ over $F$ and $a_i$ is defined in Construction \ref{constr2}.
%asbe the standard basis of $F^l$ over $F.$ 
%We write the subscript $a, 0\le a\le l-1$ in the $s$-ary form, $a=(a_n,a_{n-1},
%\dots,a_1),$ where $a_i$ is the $i$-th digit from the right (the least significant digit).
\end{cnstr}

Note that the difference between this construction and Construction \ref{constr2} is in the choice of $s$. Define $a(i,u)$ and $c_{i,a}$ in the same way as in Sect.~\ref{opt}.
  
Clearly, the code $\cC$ given by Construction \ref{constr3} is an MDS array code.
\begin{theorem}\label{ult}
The code $\cC$ given by Construction \ref{constr3} has the UER $(h,d)$-optimal repair property for any $h\le r$ and $k\le d\le n-h.$
\end{theorem}
\begin{IEEEproof}
By Theorem \ref{uer}, $\cC$ has the UER $d$-optimal repair property for any $k\le d\le n-1.$ Now we show how to optimally repair $h$ erasures. Without loss of generality, suppose that nodes $C_{{\sF}}=\{C_1,C_2,\dots,C_h\}$ are erased and we access nodes $C_{{\sR}}=\{C_{h+1},C_{h+2},\dots,C_{h+d+2t}\}$ to recover $C_{{\sF}}.$ Moreover, suppose that there are at most $t$ erroneous nodes in $C_{{\sR}}.$

To show the claim about the repair property of $\cC$, we present a scheme that repairs the codes $C_1,C_2,\dots,C_h$ one by one. More 
specifically, we first use $C_{{\sR}}$ to repair $C_1,$ then use $C_{{\sR}}\cup C_1$ to repair $C_2,$ then use $C_{{\sR}}\cup C_1\cup 
C_2$ to repair $C_3,\dots$, and finally use $C_{{\sR}} \cup C_1\cup C_2\cup\dots\cup C_{h-1}$ to repair $C_h.$ Let $s_i=i+d-k.$ 

When repairing $C_i,$ we partition the set $\{0,1,\dots,s-1\}$ into $s/s_i$ subsets 
$\cI_1^{(i)},\cI_2^{(i)},\dots,\cI_{s/s_i}^{(i)},$ 
where $\cI_j^{(i)}=\{(j-1)s_i,(j-1)s_i+1,\dots,(j-1)s_i+s_i-1\}$ for $j\in[s/s_i].$
By the proof of Theorem \ref{oldoptimal},  in order to recover $C_i$ it suffices to know the values 
$\{\sum_{u\in\cI_j^{(i)}}c_{v,a(i,u)}:a_i=0,j\in[s/s_i],v\in{\sR}\cup[i-1]\}.$ Since we have already recovered $C_1,\dots,C_{i-1},$ 
to recover $C_i$ we only need to download the set of values
   $$
   \Big\{\sum_{u\in\cI_j^{(i)}}c_{v,a(i,u)}:a_i=0,j\in[s/s_i],v\in{\sR}\Big\}
   $$ 
   from $C_{{\sR}}.$ 
Thus, in order to recover $C_{{\sF}},$ it suffices to know the values of elements in the set 
$$
\Lambda_h=\bigcup_{i=1}^h\Big\{\sum_{u\in\cI_j^{(i)}}c_{v,a(i,u)}:a_i=0,j\in[s/s_i],v\in{\sR}\Big\}.
$$
In order to determine the values of these elements, we only need to download a spanning set for $\Lambda_h$ over $F$ from $C_{{\sR}}.$

Define $\Omega_{i,v}=\{\sum_{u\in\cI_j^{(i)}}c_{v,a(i,u)}:a_i=0,j\in[s/s_i]\}$ for $i\in[h],v\in{\sR}$ and 
$\Lambda_{1,v}=\Omega_{1,v}, \Lambda_{i,v}=\Lambda_{i-1,v}\cup\Omega_{i,v}$ for $i=2,3,\dots,h,v\in{\sR}.$ 
Given $a\in\{0,1,\dots,l-1\}$ and $i\in[n],$ define the set $\Psi(a,i)=\{w:w\in[i-1],s_w|a_w\}.$
For $i=2,3,\dots,h,\,q=0,\dots,i-1$ define
    $$
\Gamma_{i,v,q}=\Big\{\sum_{u\in\cI_j^{(i)}}c_{v,a(i,u)}:a_i=0,|\Psi(a,i)|=q,j\in[s/s_i]\Big\}.
    $$
Let $B_{1,v}=\Omega_{1,v}, B_{i,v}=B_{i-1,v}\cup\Gamma_{i,v,0}, i=2,3,\dots,h.$ We use induction on $i$ to show that
$\Lambda_{i,v}\subseteq\text{span}(B_{i,v})$ for every $i\in[h]$ and $v\in{\sR}.$ 
Clearly this claim holds for $i=1.$ Now suppose that it
holds for $i=m$ and consider the case $i=m+1.$ By the induction hypothesis, $\Lambda_{m,v}\subseteq\text{span}(B_{m,v}),$ so 
it suffices to prove that 
   $$
   \Lambda_{m,v}\cup\Omega_{m+1,v}\subseteq\text{span}(\Lambda_{m,v}\cup\Gamma_{m+1,v,0}).
   $$ 
Note that $\Omega_{m+1,v}=\bigcup_{q=0}^m\Gamma_{m+1,v,q}.$ Thus we only need to prove that 
$\Gamma_{m+1,v,q}\subseteq\text{span}(\Lambda_{m,v}\cup\Gamma_{m+1,v,0})$ for all $q=0,1,\dots,m.$ We prove this claim by induction on $q.$ This claim trivially holds for $q=0.$ Now suppose that it holds for some $q\ge 1$ and consider the case $q+1.$ Given any $a$ satisfying that $a_{m+1}=0$ and $\Psi(a,m+1),|\Psi(a,m+1)|=q+1,$ suppose that $w\in\Psi(a,m+1),$ namely, $s_w|a_w,$ then $|\Psi(a(w,a_w+u),m+1)|=q$ for all $u\in[s_w-1].$
By the induction hypothesis,
$$
\sum_{u_2\in\cI_j^{(m+1)}}c_{v,a(w,m+1,a_w+u_1,u_2)}\in\text{span}(\Lambda_{m,v}\cup\Gamma_{m+1,v,0})
$$
for all $u_1\in[s_w-1]$ and $j\in[s/s_{m+1}],$ where $a(i_1,i_2,u_1,u_2)$ is obtained from $a$ by replacing $a_{i_1}$ with $u_1$ and $a_{i_2}$ with $u_2.$ Therefore,
\begin{equation}\label{eq:hard1}
\begin{aligned}
\sum_{u_1=1}^{s_w-1}\sum_{u_2\in\cI_j^{(m+1)}} &c_{v,a(w,m+1,a_w+u_1,u_2)}
\in\;\text{span\,}(\Lambda_{m,v}\cup\Gamma_{m+1,v,0}).
\end{aligned}
\end{equation}
Note that
$$
\sum_{u_1=0}^{s_w-1}c_{v,a(w,m+1,a_w+u_1,u_2)}\in\Omega_{w,v}\subseteq\Lambda_{m,v}
$$
for any $u_2\in\{0,1,\dots,s-1\}.$ As a result,
\begin{equation}\label{eq:hard2}
\sum_{u_1=0}^{s_w-1}\sum_{u_2\in\cI_j^{(m+1)}}c_{v,a(w,m+1,a_w+u_1,u_2)}\in\text{span}(\Lambda_{m,v}).
\end{equation}
Subtracting \eqref{eq:hard1} from \eqref{eq:hard2}, we obtain
    $$
\sum_{u\in\cI_j^{(m+1)}}c_{v,a(m+1,u)}\in\text{span}(\Lambda_{m,v}\cup\Gamma_{m+1,v,0})
    $$
for any $j\in[s/s_{m+1}].$ This establishes the induction step of the second induction and proves that $\Omega_{m+1,v}\subseteq\text{span}(\Lambda_{m,v}\cup\Gamma_{m+1,v,0}).$ Therefore, $\Lambda_{m+1,v}\subseteq\text{span}(B_{m+1,v})$ for any $v\in{\sR},$ and this completes the proof of the first induction.

Since $\Lambda_h=\bigcup_{v\in{\sR}}\Lambda_{h,v}$ and $\Lambda_{h,v}\subseteq\text{span}(B_{h,v})$ for every $v\in{\sR},$ 
 to recover $C_{{\sF}}$ we only need to download $\bigcup_{v\in{\sR}}B_{h,v}$ from $C_{{\sR}}.$ 
 
 Finally, we find the size of the set $B_{h,v}$ for some fixed $v\in \sR.$ 
Since $B_{i,v}=B_{i-1,v}\cup\Gamma_{i,v,0},$ we have $|B_{i,v}|\le|B_{i-1,v}|+|\Gamma_{i,v,0}|.$ 
We will prove that $|B_{i,v}|\le \frac{il}{s_i}$ by induction on $i.$ By definition $|B_{1,v}|=|\Omega_{1,v}|=\frac{l}{s_1}.$ 
 Suppose that the claim holds for $i=m$ and consider the case $i=m+1.$ It is easy to see that
\begin{align*}
|\Gamma_{m+1,v,0}|&=\frac{s_1-1}{s_1}\frac{s_2-1}{s_2}\cdots\frac{s_m-1}{s_m}\frac{l}{s_{m+1}}\\
&=\frac{d-k}{d+1-k}\frac{d+1-k}{d+2-k}\cdots\frac{d+m-1-k}{d+m-k}\frac{l}{d+m+1-k}\\
&=\frac{d-k}{d+m-k}\frac{l}{d+m+1-k}.
\end{align*}
By the induction hypothesis,
\begin{align*}
|B_{m+1,v}|&\le|B_{m,v}|+|\Gamma_{m+1,v,0}|\\
&\le\frac{ml}{d+m-k}+\frac{d-k}{d+m-k}\frac{l}{d+m+1-k}\\
&=\frac{(m+1)l}{d+m+1-k}=\frac{(m+1)l}{s_{m+1}}.
\end{align*}
This establishes the induction step and proves that  $|B_{h,v}|\le \frac{hl}{d+h-k}$ for any $v\in{\sR}.$
We obtain 
$$
\Big|\bigcup_{v\in{\sR}}B_{h,v}\Big|\le\frac{hl|{\sR}|}{d+h-k}=\frac{h(d+2t)l}{d+h-k}.
$$ The proof is complete.
\end{IEEEproof}

Since we set $A_1,A_2,\dots,A_n$ to be diagonal matrices in Construction \ref{constr3}, the encoding and repair processes involve only
 operations with $r\times r$ matrices over $F,$ and the code $\cC$ given by Construction \ref{constr3} also has the optimal update property. Moreover, by the proof of Theorem \ref{oldoptimal} and Theorem \ref{ult}, we can see that the repair process only requires operations with matrices of size no larger than $n\times n.$

\section{Optimal-repair MDS array codes with optimal access property over small fields}\label{permut}

In this section we construct an explicit family of MDS array codes with optimal access property.
As above, we rely on the general construction \eqref{eq:parityform}-\eqref{eq:power} to construct an 
$(n,k=n-r,l=r^{n-1})$ array code. 
However in this section we take $A_n$ to be the identity matrix and take $A_1,\dots,A_{n-1}$ to be permutation matrices rather
than diagonal matrices. This choice is beneficial in two ways: first, we are able to reduce the field size from $rn$ in earlier
construction to any field $F$ of size $|F|\ge n+1,$ while also obtaining the optimal access property.

\begin{cnstr}\label{constr4}
Let $F$ be a finite field of size $|F|\ge n+1$ and let $\gamma$ be its primitive element.
Let $l=r^{n-1}.$
Consider the code family given by \eqref{eq:parityform}-\eqref{eq:power}, where the matrices 
$A_1,A_2,\dots,A_n$ are 
given by
\begin{equation*}  %\label{eq:constr4}
\begin{aligned}
A_i=\sum_{a=0}^{l-1}\lambda_{i,a_i} e_a& e_{a(i,a_i\oplus 1)}^T,\; i=1,2,...,n-1,\\
A_n&=I,
\end{aligned}
\end{equation*}
where $\oplus$ denotes addition modulo $r, \lambda_{i,0}=\gamma^i$ for all $i\in[n-1]$ and $\lambda_{i,u}=1$ for all $i\in[n-1]$ and all $u\in\{1,2,\dots,r-1\}.$
Here $\{e_a:a=0,1,\dots,l-1\}$ is the standard basis of $F^l$ over $F,$ $a_i$ is the $i$-th digit from the right
in the representation of $a=(a_{n-1},a_{n-2},
\dots,a_1)$ in the $r$-ary form, and $a(i,u)$  is defined in the same way as in Sect.~\ref{opt}.
\end{cnstr}

%Clearly, the elements $1,\gamma,\gamma^2,\dots,\gamma^{n-1}$ are all distinct.

\begin{remark}
Since $\prod_{u=0}^{r-1}\lambda_{i,u}=\gamma^i,$ we have
\begin{equation}\label{eq:prodcond}
\begin{aligned}
\prod_{u=0}^{r-1}\lambda_{i,u}\neq\prod_{u=0}^{r-1}\lambda_{j,u} &\text{ for any } i,j\in[n-1],i\neq j.\\
\prod_{u=0}^{r-1}\lambda_{i,u}\neq 1 &\text{ for any } i\in[n-1].
\end{aligned}
\end{equation}
It will be clear from the proofs below that the values of $\{\lambda_{i,u}:i\in[n],u=0,1,\dots,r-1\}$ in Construction \ref{constr4}  can be assigned arbitrarily as long as \eqref{eq:prodcond} is satisfied.
\end{remark}

Clearly, for $t=0,1,\dots,r-1$ and $i\in[n-1],$ we have
    $$
A_i^t=\sum_{a=0}^{l-1}\beta_{i,a_i,t} e_a e_{a(i,a_i\oplus t)}^T,
    $$
where $\beta_{i,u,0}=1$ and $\beta_{i,u,t}=\prod_{v=u}^{u\oplus (t- 1)}\lambda_{i,v}$ for $t=1,\dots,r-1$ and $u=0,1,\dots,r-1.$

\begin{theorem}\label{opacs}
The code $\cC$ given by Construction \ref{constr4} has the optimal access property.
\end{theorem}
\begin{IEEEproof}
Let us write out the parity-check equations \eqref{eq:parityform} coordinatewise:
\begin{equation}\label{eq:scalarpermut}
\begin{aligned}
&c_{n,a}+\sum_{i=1}^{n-1} \beta_{i,a_i,t} c_{i,a(i,a_i\oplus t)}=0
\text{ for all } t=0,1,\dots,r-1 \text{ and } a=0,\dots,l-1,
\end{aligned}
\end{equation}
where $c_{i,a}$ is defined in Sect.~\ref{opt}.
First suppose we want to repair $C_i$ for some $i\in[n-1].$ We will show that we only need to access 
the values in the set $\{c_{j,a}:a_i=0\}$ from $C_j$ for every $j\neq i.$ Indeed, by \eqref{eq:scalarpermut}, we have
\begin{equation}\label{eq:optaccess}
\beta_{i,a_i,t} c_{i,a(i,a_i\oplus t)}=-c_{n,a}-\sum_{j\neq i,n} \beta_{j,a_j,t} c_{j,a(j,a_j\oplus t)}.
\end{equation}
From \eqref{eq:optaccess} we see that the values $\{c_{i,a}:a_i=t\}$ can be determined by $\{c_{j,a}:j\neq i,a_i=0\}.$ 
Since \eqref{eq:optaccess} 
holds for every $t=0,1,\dots,r-1,$ we see that for $i\in[n-1],$ $C_i$ can be determined by the values $\{c_{j,a}:j\neq i,a_i=0\}.$ 

Now consider the case when the failed node is $C_n.$ By \eqref{eq:scalarpermut}, 
we know that the values in the set $\{c_{n,a}:a_1\oplus a_2\oplus\dots\oplus a_{n-1}=r\ominus t\}$ can be determined by $\{c_{j,a}:j\neq n,a_1\oplus 
a_2\oplus\dots\oplus a_{n-1}=0\},$ where $\ominus$ denotes subtraction modulo $r.$ Since \eqref{eq:scalarpermut} holds for every 
$t=0,1,\dots,r-1,$ 
we conclude that $C_n$ can be determined by $\{c_{j,a}:j\neq n,a_1\oplus a_2\oplus\dots\oplus a_{n-1}=0\}.$
This completes the proof.
\end{IEEEproof}

Our next task is to establish the MDS property of $\cC.$ The code $\cC$ is MDS if and only if every $r\times r$ block submatrix of
$$
\left[ \begin{array}{cccc} I & I & \dots & I\\
A_1 & A_2 & \dots & A_n\\
\vdots & \vdots & \vdots & \vdots\\
A_1^{r-1} & A_2^{r-1} & \dots & A_n^{r-1} \end{array} \right]
$$
is invertible. A criterion for this is given in the following lemma.
\begin{lemma}[Block Vandermonde matrix]\label{vand}
Let $B_1,\dots,B_r$ be $l\times l$ matrices such that $B_iB_j=B_jB_i$ for all $i,j\in[r].$ The matrix
$$
M_r=\left[ \begin{array}{cccc} I & I & \dots & I\\
B_1 & B_2 & \dots & B_r\\
\vdots & \vdots & \vdots & \vdots\\
B_1^{r-1} & B_2^{r-1} & \dots & B_r^{r-1}\\ \end{array} \right]
$$
is invertible if and only if $B_i-B_j$ is invertible for all $i\neq j.$
\end{lemma}
\begin{IEEEproof}
Suppose that $B_i-B_j$ is invertible for any $i,j\in[r],i\neq j.$  Clearly the claim holds for $r=2.$ 
Now suppose that it holds for $r=s.$ Consider the matrix
$$
M_{s+1}=\left[ \begin{array}{cccc} I & I & \dots & I\\
B_1 & B_2 & \dots & B_{s+1}\\
\vdots & \vdots & \vdots & \vdots\\
B_1^s & B_2^s & \dots & B_{s+1}^s\\ \end{array} \right].
$$
For $i=s,s-1,\dots,1,$ multiply the $i$-th ``row'' on the left by $B_1$ and subtract from the $(i+1)$-th row.  
Note that these operations do not change the 
rank of $M_{s+1}$ since they leave its row space unchanged. Next, subtract the first column from all the other 
columns (this clears the first row without changing the column space and hence the rank of $M_{s+1}$) to obtain
\begin{align*}
M_{s+1}'=\left[ \begin{array}{cccc} I & 0 & \dots & 0\\
0 & B_2-B_1 & \dots & B_{s+1}-B_1\\
\vdots & \vdots & \vdots & \vdots\\
0 & (B_2-B_1)B_2^{s-1} & \dots & (B_{s+1}-B_1)B_{s+1}^{s-1}\\ \end{array} \right]\\
{=}\left[ \begin{array}{cccc} I & 0 & \dots & 0\\
0 & B_2-B_1 & \dots & B_{s+1}-B_1\\
\vdots & \vdots & \vdots & \vdots\\
0 & B_2^{s-1}(B_2-B_1) & \dots & B_{s+1}^{s-1}(B_{s+1}-B_1)\\ \end{array} \right]
\end{align*}
(here we relied on the commuting condition $B_iB_j=B_jB_i$). 
Since the matrices $B_i-B_1,i=2,\dots,s+1$ are invertible, we can multiply 
the $i$-th column on the right by $(B_i-B_1)^{-1}$ without changing the rank. We conclude that the matrix
$$
M_{s+1}^{''}=\left[ \begin{array}{cccc} I & 0 & \dots & 0\\
0 & I & \dots & I\\
\vdots & \vdots & \vdots & \vdots\\
0 & B_2^{s-1} & \dots & B_{s+1}^{s-1}\\ \end{array} \right]
$$
has the same rank as $M_{s+1}.$ By the induction hypothesis $M_{s+1}^{''}$ is invertible, and so is $M_{s+1}.$ This completes the induction step.

Conversely, suppose that $M_r$ is invertible. For $r=2$ this implies that the matrix $B_1-B_2$ is invertible.  Now assume that the claim holds for $r=s$ and consider the case $r=s+1.$ Since $M_{s+1}$ is invertible, $M'_{s+1}$ is also invertible. Assume that $B_i-B_1$ is singular for some $i\in\{2,3,\dots,s+1\},$ then
$$
\left[ \begin{array}{c}  0\\
 B_i-B_1\\
 \vdots\\
B_i^{s-1} (B_i-B_1)\\ \end{array} \right]
=\left[ \begin{array}{c}  0\\
 I\\
 \vdots\\
B_i^{s-1} \\ \end{array} \right] (B_i-B_1)
$$
contains linearly dependent columns, and thus $M'_{s+1}$ is singular, contradiction. Thus $B_i-B_1$ is invertible for any $i\neq 1.$ Consequently $M_{s+1}^{''}$ is invertible. By the induction assumption we conclude that $B_i-B_j$ is invertible for any $i,j\in[s+1],i\neq j.$ 
%This completes the proof of the inductive step and the converse part.
\end{IEEEproof}

\begin{theorem}\label{3MDS}
The code $\cC$ given by Construction \ref{constr4} is an MDS array code.
\end{theorem}
\begin{IEEEproof} On account of
Lemma \ref{vand}, the claim will follow if we prove that for any $i\neq j, A_iA_j=A_jA_i$ and that the matrices $A_i-A_j$ are invertible. The matrix $A_n=I,$ so we need to verify that for any $i,j\in[n-1],i\neq j,$
$$
A_iA_j=A_jA_i=\sum_{a=0}^{l-1}\lambda_{i,a_i}\lambda_{j,a_j}e_a e_{a(i,j,a_i\oplus 1,a_j\oplus 1)}^T,
$$
where $a(i,j,u,v)$ is obtained from $a$ by replacing $a_i$ with $u$ and $a_j$ with $v.$
%and $b_m=a_m$ for all $m\neq i,j.$ 
This establishes the commuting part.

Now suppose that $A_ix=A_jx$ for some $i,j\in[n-1],i\neq j$ and some vector $x\in F^l.$ Let $x=\sum_{a=0}^{l-1}x_ae_a,$ where $x_a\in F.$ Then
\begin{align*}
A_ix=\sum_{a=0}^{l-1}\lambda_{i,a_i}x_{a(i,a_i\oplus 1)}e_a,\\
A_jx=\sum_{a=0}^{l-1}\lambda_{j,a_j}x_{a(j,a_j\oplus 1)}e_a.
\end{align*}
Therefore,
\begin{equation}\label{eq:temp}
\lambda_{i,a_i}x_{a(i,a_i\oplus 1)}=\lambda_{j,a_j}x_{a(j,a_j\oplus 1)}
\end{equation}
for every $a=0,1,\dots,l-1.$ Since $\lambda_{i,u}\neq 0$ for all $i\in[n-1]$ and all $u=0,1,\dots,r-1,$ we can write \eqref{eq:temp} as
\begin{equation}\label{eq:invertible}
x_a=\frac{\lambda_{j,a_j}}{\lambda_{i,a_i\ominus 1}}x_{a(i,j,a_i\ominus 1,a_j\oplus 1)}.
\end{equation}
Repeating this step, we obtain
% Replacing $a$ in \eqref{eq:invertible} with $a(i,j,a_i\ominus 1,a_j\oplus 1),a(i,j,a_i\ominus 2,a_j\oplus 2),\dots,a(i,j,a_i\ominus (r-1),a_j\oplus (r-1))$ respectively, we get 
\begin{align*}
x_a&=\frac{\lambda_{j,a_j}}{\lambda_{i,a_i\ominus 1}}x_{a(i,j,a_i\ominus 1,a_j\oplus 1)}\\
&=\frac{\lambda_{j,a_j}}{\lambda_{i,a_i\ominus 1}}\frac{\lambda_{j,a_j\oplus 1}}{\lambda_{i,a_i\ominus 2}}x_{a(i,j,a_i\ominus 2,a_j\oplus 2)}\\
&=\dots=\frac{\lambda_{j,a_j}}{\lambda_{i,a_i\ominus 1}}\frac{\lambda_{j,a_j\oplus 1}}{\lambda_{i,a_i\ominus 2}}\ldots\frac{\lambda_{j,a_j\oplus(r-1)}}{\lambda_{i,a_i\ominus r}} x_{a(i,j,a_i\ominus r,a_j\oplus r)}\\
&=\frac{\prod_{u=0}^{r-1}\lambda_{j,u}}{\prod_{u=0}^{r-1}\lambda_{i,u}}x_a
\end{align*}
for every $a=0,1,\dots,l-1.$ By \eqref{eq:prodcond},  $x_a=0$ for all $a=0,\dots,l-1.$ Thus $(A_i-A_j)x=0$ implies that $x=0.$ We conclude 
that the matrices $A_i-A_j$ are invertible for all $i,j\in[n-1],i\neq j.$

Now suppose that $A_ix=A_nx=x$ for some $i\in[n-1]$ and some vector $x\in F^l.$ Then we have
\begin{equation}\label{eq:an}
\lambda_{i,a_i}x_{a(i,a_i\oplus 1)}=x_a.
\end{equation}
Thus
\begin{align*}
&x_a=\lambda_{i,a_i}x_{a(i,a_i\oplus 1)}=\lambda_{i,a_i}\lambda_{i,a_i\oplus 1}x_{a(i,a_i\oplus 2)}\\
&=\dots=\lambda_{i,a_i}\lambda_{i,a_i\oplus 1}\cdots \lambda_{i,a_i\oplus (r-1)}x_{a(i,a_i\oplus r)}\\
&=\big(\prod_{u=0}^{r-1}\lambda_{i,u}\big)x_a.
\end{align*}
By \eqref{eq:prodcond},  $x_a=0$ for all $a=0,\dots,l-1.$ So $x=0$ and $\ker(A_i-A_n)=0,$ or, in other words the matrix $A_i-A_n$ is invertible for all $i\in[n-1].$ %We conclude that the matrices $A_i-A_j$ are invertible for any $i,j\in[n],i\neq j.$ 
This completes the proof.
\end{IEEEproof}

\subsection{Complexity of encoding and decoding}
The encoding and decoding procedures solve the same problem, namely, determining
the values of $r$ nodes from the known values of $k$ nodes. Without loss of generality, 
suppose that we want to 
determine $C_{k+1},\dots,C_{k+r}$ from $C_1,C_2,\dots,C_k.$ Note that \eqref{eq:scalarpermut} contains $rl$ equations and we have $rl$ 
unknown elements here. Since the code is MDS, the unknown values are uniquely determined. 

Now let us show that instead of inverting an $rl\times rl
$ matrix, we only need to invert matrices of size $r^{r+1}\times r^{r+1}.$ 
Observe that, given any $b_1,b_2,\dots,b_k\in \{0,1,\dots,r-1\},$ the $r^{r+1}$ unknown elements 
$\{c_{i,a}:i=k+1,k+2,\dots,k+r,a_1=b_1,a_2=b_2,\dots,a_k=b_k\}$ appear in exactly 
$r^{r+1}$ equations in \eqref{eq:scalarpermut}, and these $r^{r+1}$ equations only contain these $r^{r+1}$ unknown elements. For this 
reason, we can find these $r^{r+1}$ unknown elements by inverting an $r^{r+1}\times r^{r+1}$ matrix. 
%Thus we showed that in both 
%encoding and decoding procedures, the calculation only involves the inversion of $r^{r+1}\times r^{r+1}$ matrices.

\section{Explicit MDS array codes with the UER $d$-optimal access property}\label{2d}

\begin{cnstr}\label{constr5} Let $F$ be a finite field of size $|F|\ge n+1$ and let $\gamma$ be a primitive element in $F.$ Let $s=d+k-1$ and $l=s^n.$ 
Consider the code given by \eqref{eq:parityform}-\eqref{eq:power}, where
the matrices $A_1,A_2,\dots,A_n$ are given by
$$
A_i=\sum_{a=0}^{l-1}\lambda_{i,a_i}e_a e_{a(i,a_i\oplus 1)}^T , \; i=1,\dots,n,
$$
where $\oplus$ denotes addition modulo $s, \lambda_{i,0}=\gamma^i$ for all $i\in[n]$ and $\lambda_{i,u}=1$ for all $i\in[n]$ and all $u\in[s-1].$ 
Here $\{e_a:a=0,1,\dots,l-1\}$ is the standard basis of $F^l$ over $F,$ $a_i$ is the  $i$-th digit from the right in the
representation of $a=(a_n,a_{n-1},
\dots,a_1)$ in the $s$-ary form, and $a(i,u)$ is defined in the same way as in Sect.~\ref{opt}.
\end{cnstr}

\begin{theorem}
The code $\cC$ given by Construction \ref{constr5} is an MDS array code.
\end{theorem}
\begin{IEEEproof} 
Paralleling the proof of Theorem \ref{3MDS}, we can show that  for any $i\neq j,$ $A_iA_j=A_jA_i$ and that the matrix $A_i-A_j$ is invertible. Thus $\cC$ is an MDS array code.
\end{IEEEproof}

\begin{theorem}
The code $\cC$ given by Construction \ref{constr5} has the UER $d$-optimal access property. 
\end{theorem}
\begin{IEEEproof} 
Suppose we want to repair $C_i.$ By an argument similar to the proof of Theorem \ref{opacs}, 
we only need to know the values $\{c_{j,a}:a_i=0\}$ from $C_j$ for every $j\neq i.$ 
Define a function $g:\{0,1,\dots,l/s-1\}\to\{0,1,\dots,l-1\}$ as $g(a)=(a_{n-1},a_{n-2},\dots,a_i,0,a_{i-1},a_{i-2},\dots,a_1),$ 
where $a$ is an element in $\{0,1,\dots,l/s-1\}$ with the $s$-ary expansion $(0,a_{n-1},a_{n-2},\dots,a_1).$ 
Define the column vector $C_j^{(i)}\in F^{l/s}$ as $C_j^{(i)}=(c_{j,g(0)},c_{j,g(1)}, \dots, c_{j,g(l/s-1)})^T$ for 
all $j\neq i.$ In order to prove the theorem, we only need to prove that 
$(C_1^{(i)},C_2^{(i)},\dots,C_{i-1}^{(i)},C_{i+1}^{(i)},C_{i+2}^{(i)},\dots,C_n^{(i)})$ forms an $(n-1,d,s^{n-1})$ MDS array code.

Notice that $A_j^s=\sum_{a=0}^{l-1}(\prod_{u=0}^{s-1}\lambda_{j,u})e_a e_{a(j,a_j\oplus s)}^T=\gamma^j I$ for all $j\in[n].$ Note also that
$$
\sum_{j=1}^n A_j^m C_j=0, \quad
\sum_{j=1}^n A_j^{m+s} C_j=0
$$
for all $m=0,1,\dots,r-s-1.$ Multiplying the first equation by $\gamma^i$ and then subtracting it from the second one, we obtain
\begin{equation}\label{eq:atob}
\sum_{j\neq i} (\gamma^j-\gamma^i)A_j^mC_j=0,\quad m=0,1,\dots,r-s-1.
\end{equation}
Let $e_0^{(l/s)},e_1^{(l/s)},\dots,e_{l/s-1}^{(l/s)}$ be the standard basis vectors of $F^{l/s}$ over $F.$ Define $l/s\times l/s$ matrix 
\begin{align*}
B_j=\sum_{a=0}^{l/s-1}\lambda_{j,a_j}e_a^{(l/s)} (e_{a(j,a_j\oplus 1)}^{(l/s)})^T \text{ for } j\in[i-1],\\
B_j=\sum_{a=0}^{l/s-1}\lambda_{j+1,a_j}e_a^{(l/s)} (e_{a(j,a_j\oplus 1)}^{(l/s)})^T \text{ for } i\le j<n,
\end{align*}
 It is easy to see that \eqref{eq:atob} implies
\begin{align*}
\sum_{j=1}^{i-1}(\gamma^j-\gamma^i)B_j^m &C_j^{(i)}+\sum_{j=i}^{n-1}(\gamma^{j+1}-\gamma^i)B_j^mC_{j+1}^{(i)}=0\\
\text{ for all } &m=0,1,\dots,r-s-1.
\end{align*}
As in the proof of Theorem \ref{3MDS},  we can show that  for any $j_1,j_2\in[n-1],j_1\neq j_2,$ $B_{j_1}B_{j_2}=B_{j_2}B_{j_1}$ and that the matrix $B_{j_1}-B_{j_2}$ is invertible. Moreover, $r-s=n-1-d.$ Thus
$$
\left[ \begin{array}{cccc} I & I & \dots & I\\
B_1 & B_2 & \dots & B_{n-1}\\
\vdots & \vdots & \vdots & \vdots\\
B_1^{r-s-1} & B_2^{r-s-1} & \dots & B_{n-1}^{r-s-1}\\ \end{array} \right]
$$
is a parity-check matrix of an $(n-1,d,s^{n-1})$ MDS array code. Multiplying each block column with a nonzero constant 
does not change the MDS property.
As a result, the set of vectors $(C_1^{(i)},C_2^{(i)},\dots,C_{i-1}^{(i)},C_{i+1}^{(i)},C_{i+2}^{(i)},\dots,C_n^{(i)})$ forms an $(n-1,d,s^{n-1})$ MDS array code. Therefore, if we access any $d+2t$ out of $n-1$ vectors in the set  
$(C_1^{(i)},C_2^{(i)},\dots, C_{i-1}^{(i)},C_{i+1}^{(i)},C_{i+2}^{(i)},\dots,C_n^{(i)}),$ we will be able to recover the whole set and further recover $C_i$ as long as the number of erroneous nodes among the helper nodes is not greater than $t.$ This completes the proof.
\end{IEEEproof}

\section{An MDS array code family with the UER $d$-optimal access property for several values of $d$ simultaneously}\label{2md}
In this section we present a simple extension of the code family in Construction \ref{constr5} which gives 
MDS array codes with the UER $d$-optimal access property for several values of $d$ simultaneously. 
More specifically, given any positive integers $n,k,m,d_1,d_2,\dots,d_m$ such that $k\le d_1,\dots,d_m<n,$ we will show that by replacing $s$ in Construction \ref{constr5} with the value
$$
s=\text{lcm}(d_1+1-k,d_2+1-k,\dots,d_m+1-k)
$$
we obtain an $(n,k,l=s^n)$  MDS array code $\cC$ with the UER $d_i$-optimal access property for all $i=1,\dots,m$ simultaneously.

On account of Theorem \ref{3MDS}, we already know that $\cC$ is an MDS array code. It remains to show that it has the UER $d_i$-optimal access property for any $i\in[m].$

\begin{theorem}\label{univac}
The code $\cC$ has the UER $d_i$-optimal access property for any $i\in[m].$
\end{theorem}
\begin{IEEEproof}
Given any $i\in[m],$ we show that $\cC$ has the UER $d_i$-optimal access property. Let $s_i=d_i+1-k.$ 

Without loss of generality, we only consider the case of repairing $C_n.$ By an argument similar to the proof of Theorem \ref{opacs}, we only need to know the values $\{c_{j,a}:a_n=0,s_i,2s_i,\dots,(s/s_i-1)s_i\}$ from $C_j$ for every $j\in[n-1].$ 
Define a function 
   \begin{align*}
   g:\{0,1,\dots,l/s_i-1\}&\to\{0,1,\dots,l-1\}\\
   a\hspace*{.5in} &\mapsto (s_ia_n,a_{n-1},a_{n-2},\dots,a_1),
   \end{align*}
    where $a$ is an element in $\{0,1,\dots,l/s_i-1\}$ with $s$-ary expansion $(a_n,a_{n-1},a_{n-2},\dots,a_1).$
Define the column vector $C_j^{(n)}\in F^{l/s_i}$ as $C_j^{(n)}=(c_{j,g(0)}, c_{j,g(1)},\dots, c_{j,g(l/s_i-1)})^T$ for all $j\in[n-1].$ As mentioned above, $\{C_j^{(n)}:j\in[n-1]\}$ contains the information we need to recover $C_n.$ Let us prove that $(C_1^{(n)},C_2^{(n)},\dots,C_{n-1}^{(n)})$ forms an $(n-1,d,l/s_i)$ MDS array code.

Observe that
$$
\sum_{j=1}^n A_j^m C_j=0, \quad
\sum_{j=1}^n A_j^{m+s_i} C_j=0
$$
for all $m=0,1,\dots,r-s_i-1.$ Multiplying the first equation on the left by $A_n^{s_i}$ and then subtracting it from the second one, we obtain
\begin{equation}\label{eq:astob}
\sum_{j=1}^{n-1} (A_j^{s_i}-A_n^{s_i})A_j^mC_j=0,\quad m=0,1,\dots,r-s_i-1.
\end{equation}
Let $e_0^{(l/s_i)},e_1^{(l/s_i)},\dots,e_{l/s_i-1}^{(l/s_i)}$ be the standard basis vectors of $F^{l/s_i}$ over $F.$
Define $l/s_i\times l/s_i$ matrices
\begin{align*}
B_j&=\sum_{a=0}^{l/s_i-1}\lambda_{j,a_j}e_a^{(l/s_i)} (e_{a(j,a_j\oplus 1)}^{(l/s_i)})^T \text{ for } j\in[n-1],\\
B_n&=\sum_{a=0}^{l/s_i-1}\Big(\prod_{q=s_ia_n}^{s_ia_n+s_i-1}\lambda_{n,q}\Big)e_a^{(l/s_i)} (e_{a(n,(s_ia_n\oplus s_i)/s_i)}^{(l/s_i)})^T.
\end{align*}
It is easy to see that \eqref{eq:astob} implies the equality
$$
\sum_{j=1}^{n-1} (B_j^{s_i}-B_n)B_j^mC_j^{(n)}=0,\quad m=0,1,\dots,r-s_i-1.
$$
Now suppose that $B_j^{s_i}x=B_nx$ for some $j\in[n-1]$ and some vector $x\in F^{l/s_i}.$ Let $x=\sum_{a=0}^{l/s_i-1}x_ae_a^{(l/s_i)},$ where $x_a\in F.$ Then
\begin{align*}
B_j^{s_i}x&=\sum_{a=0}^{l/s_i-1}\Big(\prod_{q=a_j}^{a_j\oplus(s_i-1)}\lambda_{j,q}\Big)x_{a(j,a_j\oplus s_i)}e_a^{(l/s_i)},\\
B_nx&=\sum_{a=0}^{l/s_i-1}\Big(\prod_{q=s_ia_n}^{s_ia_n+s_i-1}\lambda_{n,q}\Big)x_{a(n,(s_ia_n\oplus s_i)/s_i)}e_a^{(l/s_i)}.
\end{align*}
Therefore,
\begin{equation}\label{eq:conti}
\begin{aligned}
&\Big(\prod_{q=a_j}^{a_j\oplus(s_i-1)}\lambda_{j,q}\Big)x_{a(j,a_j\oplus s_i)}
%\\&
=\Big(\prod_{q=s_ia_n}^{s_ia_n+s_i-1}\lambda_{n,q}\Big)x_{a(n,(s_ia_n\oplus s_i)/s_i)}
\end{aligned}
\end{equation}
for every $a=0,1,\dots,l/s_i-1.$ Let us rewrite \eqref{eq:conti} as
$$
x_a=\frac{\prod_{q=s_ia_n}^{s_ia_n+s_i-1}\lambda_{n,q}}{\prod_{q=a_j\ominus s_i}^{a_j\ominus 1}\lambda_{j,q}}x_{a(j,n,a_j\ominus s_i,(s_ia_n\oplus s_i)/s_i)}.
$$
Repeating this step, we obtain
% Replacing $a$ in \eqref{eq:invertible} with $a(i,j,a_i\ominus 1,a_j\oplus 1),a(i,j,a_i\ominus 2,a_j\oplus 2),\dots,a(i,j,a_i\ominus (r-1),a_j\oplus (r-1))$ respectively, we get 
\begin{align*}
x_a&=\frac{\prod_{q=s_ia_n}^{s_ia_n\oplus(s-1)}\lambda_{n,q}}{\prod_{q=a_j\ominus s}^{a_j\ominus 1}\lambda_{j,q}}x_{a(j,n,a_j\ominus s,(s_ia_n\oplus s)/s_i)}\\
&=\frac{\prod_{u=0}^{s-1}\lambda_{n,u}}{\prod_{u=0}^{s-1}\lambda_{j,u}}x_a\\
&=\frac{\gamma^n}{\gamma^j}x_a
\end{align*}
for every $a=0,1,\dots,l-1.$ Since $\gamma^j\neq \gamma^n$ for any $j\in[n-1],$ we have $x_a=0$ for all $a=0,\dots,l-1.$ Thus $(B_j^{s_i}-B_n)x=0$ implies that $x=0.$ We conclude 
that the matrix $B_j^{s_i}-B_n$ is invertible for all $j\in[n-1].$

Following the proof of Theorem \ref{3MDS}, we can show that  
   $$
\left[ \begin{array}{cccc} I & I & \dots & I\\
B_1 & B_2 & \dots & B_{n-1}\\
\vdots & \vdots & \vdots & \vdots\\
B_1^{r-s_i-1} & B_2^{r-s_i-1} & \dots & B_{n-1}^{r-s_i-1}\\ \end{array} \right]
  $$
is the parity-check matrix of an $(n-1,d_i,l/s_i)$ MDS array code. Multiplying each block column with an invertible matrix does not change the MDS property.
As a result,  $(C_1^{(n)},C_2^{(n)},\dots,C_{n-1}^{(n)})$ forms an $(n-1,d_i,l/s_i)$ MDS array code. Therefore, if we access any 
$d_i+2t$ out of $n-1$ vectors in the set  $(C_1^{(n)},C_2^{(n)},\dots,C_{n-1}^{(n)}),$ we will be able to recover 
the whole set and further recover $C_n$ as long as the number of erroneous nodes among the helper nodes is not greater than $t.$ This 
completes the proof.
\end{IEEEproof}

\section{Explicit  MDS array codes with the UER $(h,d)$-optimal access property for all $h\le r$ and $k\le d\le n-h$ simultaneously}\label{2ult}

Given integers $n$ and $r,$
we construct a family of $(n,k=n-r,l)$ MDS array codes 
with the UER $(h,d)$-optimal access property for all $h\le r$ and $k\le d\le n-h$ simultaneously.

\begin{cnstr}\label{constr6}
Let $F$ be a finite field of size $|F|\ge n+1$ and let $\gamma$ be its primitive element. 
Let $s=\text{lcm}(1,2,\dots,r)$ and $l=s^n.$
Consider the code family given by \eqref{eq:parityform}-\eqref{eq:power}, where the matrices $A_i$ are given
by
$$
A_i=\sum_{a=0}^{l-1}\lambda_{i,a_i}e_a e_{a(i,a_i\oplus 1)}^T, \; i=1,\dots,n,
$$
where $\oplus$ denotes addition modulo $s, \lambda_{i,0}=\gamma^i$ for all $i\in[n]$ and $\lambda_{i,u}=1$ for all $i\in[n]$ and all $u\in[s-1].$ Here $\{e_a:a=0,1,\dots,l-1\}$ is the standard basis of $F^l$ over $F,$ $a_i$ is the
  $i$-th digit from the right in the representation of $a=(a_n,a_{n-1},
\dots,a_1)$ in the $s$-ary form, and $a(i,u)$ is defined in the same way as in Sect.~\ref{opt}.
\end{cnstr}

Note that the difference between this construction and Construction \ref{constr5} is in the choice of $s$.

Clearly, the code $\cC$ given by Construction \ref{constr6} is an MDS array code.
\begin{theorem}\label{help}
The code $\cC$ given by Construction \ref{constr6} has the UER $(h,d)$-optimal access property for all $h\le r$ and $k\le d\le n-h$ simultaneously.
\end{theorem}
\begin{IEEEproof}
By Theorem \ref{univac}, $\cC$ has the UER $d$-optimal access property for any $k\le d\le n-1.$ Now we show how to optimally repair $h$ erasures. Without loss of generality, suppose that nodes $C_{{\sF}}=\{C_1,C_2,\dots,C_h\}$ are erased and we access nodes $C_{{\sR}}=\{C_{h+1},C_{h+2},\dots,C_{h+d+2t}\}$ to recover $C_{{\sF}}.$ Moreover, suppose that there are at most $t$ erroneous nodes in $C_{{\sR}}.$

As before, we repair $C_1,C_2,\dots,C_h$ one by one. More specifically, we first use $C_{{\sR}}$ to repair $C_1,$ then  $C_{{\sR}}\cup C_1$ to repair $C_2,$ then $C_{{\sR}}\cup C_1\cup C_2$ to repair $C_3,\dots$, and finally we use $C_{{\sR}}\cup C_1\cup C_2\cup\dots\cup C_{h-1}$ to repair $C_h.$ Let $s_i=i+d-k.$ 
When repairing $C_i,$ 
according to the proof of Theorem \ref{univac}, we only need to know the values
$\{c_{v,a}:a_i=0,s_i,2s_i,\dots,(s/s_i-1)s_i,v\in{\sR}\bigcup[i-1]\}.$ Since we have already recovered $C_1,\dots,C_{i-1},$ we  need to access only the values $\{c_{v,a}:a_i=0,s_i,2s_i,\dots,(s/s_i-1)s_i,v\in{\sR}\}$ from $C_{{\sR}}$ to recover $C_i.$
Thus in order to recover $C_{{\sF}},$ we need to access the set of elements $\Lambda_h=\bigcup_{i=1}^h\{c_{v,a}:a_i=0,s_i,2s_i,\dots,(s/s_i-1)s_i,v\in{\sR}\}.$

Consider the set 
  $$\Omega_{j,v}=\{c_{v,a}:a_j=0,s_j,2s_j,\dots,(s/s_j-1)s_j\}$$
   for $j\in[h]$ and $v\in{\sR}.$ 
   Let $\Lambda_{1,v}=\Omega_{1,v}, \Lambda_{j+1,v}=\Lambda_{j,v}\cup\Omega_{j+1,v}, j=1,\dots,h-1.$
We prove by induction on $j$ that $|\Lambda_{j,v}|=\frac{jl}{j+d-k}.$ Clearly this is true for $j=1.$ 
Suppose that the claim is true for $j=m$ and consider the case $j=m+1.$ 
By definition, we have 
$|\Lambda_{m+1,v}|=|\Lambda_{m,v}|+|\Omega_{m+1,v}|-|\Lambda_{m,v}\cap\Omega_{m+1,v}|.$ 
By the induction hypothesis, $|\Lambda_{m,v}|=\frac{ml}{m+d-k}.$ Therefore,
   $$
|\Lambda_{m,v}\bigcap\Omega_{m+1,v}|=\frac{ml}{m+d-k}\frac{1}{m+1+d-k}.
   $$
Thus
\begin{align*}
|\Lambda_{m+1,v}|&=\frac{ml}{m+d-k}
%\\&
+\frac{l}{m+1+d-k}-\frac{ml}{m+d-k}\frac{1}{m+1+d-k}\\
&=\frac{(m+1)l}{m+1+d-k}.
\end{align*}
This concludes the induction step and proves that $|\Lambda_{h,v}|= \frac{hl}{d+h-k}$ for any $v\in{\sR}.$ Thus $|\Lambda_h|=|{\sR}||\Lambda_{h,v}|=\frac{h(d+2t)l}{d+h-k}.$ The proof is complete.
\end{IEEEproof}

\section{Generalized Reed-Solomon Array codes and $d$-optimal repair property}\label{GRSA}
In this section we construct a family of MDS array codes with the $d$-optimal repair property that requires a
smaller underlying field size compared to Construction \ref{constr2} and a 
smaller $l$ compared to both Construction \ref{constr2} and Construction \ref{constr5}. The construction forms an extension of  
Construction \ref{constr4}. As a first step, we introduce a new class of MDS array 
codes.

\subsection{Generalized Reed-Solomon Array Codes}
\begin{definition}\label{defgrsa}
Let $\sA=\{A_1,A_2,\dots,A_n\}$ be a set of $l\times l$ matrices over $F$ such that 
$A_iA_j=A_jA_i$ for all $i,j\in[n]$ and the matrices $A_i-A_j$ are invertible for all $i\neq j.$ Let $\sV=\{V_1,V_2,\dots,V_n\}$ be a set of $l\times l$ 
invertible matrices with entries in $F$ such that $A_iV_j=V_jA_i$ for any $i,j\in[n].$
A \emph{Generalized Reed-Solomon array code} $\text{\rm GRSA}(n,k,\sA,\sV)$ is defined as the $(n,k,l)$ array code with 
the generator matrix
$$
G=\left[ \begin{array}{cccc} V_1 & V_2 & \dots & V_n\\
A_1V_1 & A_2V_2 & \dots & A_nV_n\\
A_1^2V_1 & A_2^2V_2 & \dots & A_n^2V_n\\
\vdots & \vdots & \vdots & \vdots\\
A_1^{k-1}V_1 & A_2^{k-1}V_2 & \dots & A_n^{k-1}V_n \end{array} \right].
$$
More specifically,
\begin{equation}\label{eq:gen}
\begin{aligned}
\text{\rm GRSA}&(n,k,\sA,\sV)=\{(C_1,C_2,\dots,C_n):[C_1^T C_2^T \dots C_n^T]\\
=&[M_1^T M_2^T \dots M_k^T]G \text{ for some } M_1,\dots,M_k\in F^l\}.
\end{aligned}
\end{equation}
If $V_1=V_2=\dots=V_n=I,$ then we call this code a Reed-Solomon array code and denote it as $\text{\rm RSA}(n,k,\sA).$
\end{definition}

Lemma \ref{vand} implies that GRSA codes have the MDS property.
We need a description of its dual code, which is analogous to the scalar case (Theorem 10.4 in \cite[p.304]{Macwilliams77}).

\begin{theorem}\label{dualgrsa}
Given a Generalized Reed-Solomon array code $\text{\rm GRSA}(n,k=n-r,\sA,\sV)$ with $\sA$ and $\sV$ satisfying the conditions 
in Def.~\ref{defgrsa}, there is a set $\sW=\{W_1,W_2,\dots,W_n\}$ of $l\times l$ 
invertible matrices such that $A_iW_j=W_jA_i$ for any $i,j\in[n],$ and
\begin{equation}\label{eq:parity}
\begin{aligned}
&\text{\rm GRSA}(n,k,\sA,\sV)
%\\&
=\{(C_1,C_2,\dots,C_n):H[C_1^T C_2^T \dots C_n^T]^T=0\},
\end{aligned}
\end{equation}
where
$$
H=\left[ \begin{array}{cccc} W_1 & W_2 & \dots & W_n\\
A_1W_1 & A_2W_2 & \dots & A_nW_n\\
A_1^2W_1 & A_2^2W_2 & \dots & A_n^2W_n\\
\vdots & \vdots & \vdots & \vdots\\
A_1^{r-1}W_1 & A_2^{r-1}W_2 & \dots & A_n^{r-1}W_n \end{array} \right].
$$
\end{theorem}
This theorem follows from the following lemma.
\begin{lemma}
Let $A_1,A_2,\dots,A_n$ be $l\times l$ matrices with entries in $F$ such that $A_iA_j=A_jA_i$ for all $i,j\in[n]$ and the matrices $A_i-A_j$ 
are invertible for all $i\neq j.$ The following equation holds:
\begin{equation}\label{eq:identity}
\left[ \begin{array}{cccc} I & I & \dots & I\\
A_1 & A_2 & \dots & A_n\\
A_1^2 & A_2^2 & \dots & A_n^2\\
\vdots & \vdots & \vdots & \vdots\\
A_1^{n-2} & A_2^{n-2} & \dots & A_n^{n-2} \end{array} \right]
\left[ \begin{array}{c} B_1\\
B_2\\
\vdots \\
B_n \end{array} \right]=0,
\end{equation}
where $B_i=(\prod_{j\neq i}(A_j-A_i))^{-1}, i=1,\dots,n$
\end{lemma}
\begin{IEEEproof}
Note that the theory of determinants as well as Cramer's rule extend
 with no extra effort over arbitrary commutative rings with identity 
\cite[Sect. 11.4]{dummit91}. Let $R$ be a commutative ring containing 
$I,A_1,A_2,\dots,A_n.$ Given a square block matrix partitioned into 
$l\times l$ square submatrices in $R,$ we view it as a square matrix with 
entries in $R$ and define the determinant accordingly. Clearly, 
for any $i\in[n]$ and any nonnegative integer $t,$ $A_i^t\in R.$ We use Cramer's rule 
to solve the 
following equation
{$$
\left[ \begin{array}{cccc} I & I & \dots & I\\
A_1 & A_2 & \dots & A_{n-1}\\
A_1^2 & A_2^2 & \dots & A_{n-1}^2\\
\vdots & \vdots & \vdots & \vdots\\
A_1^{n-2} & A_2^{n-2} & \dots & A_{n-1}^{n-2} \end{array} \right]
\left[ \begin{array}{c} X_1\\
X_2\\
\vdots \\
X_{n-1} \end{array} \right]=
-\left[ \begin{array}{c} I\\
A_n\\
A_n^2\\
\vdots \\
A_n^{n-2} \end{array} \right]
$$}
The expression for the Vandermonde determinant also holds in commutative rings, and we can easily find that
$X_i=(\prod_{j\neq n}(A_j-A_n))(\prod_{j\neq i}(A_j-A_i))^{-1}.$ Moving the right-hand side of the equation above to the left and then
multiplying on the right by $B_n$, we obtain \eqref{eq:identity}. \end{IEEEproof}

\begin{IEEEproof}[Proof of Theorem \ref{dualgrsa}]
Set $W_i=V_i^{-1}(\prod_{j\neq i}(A_j-A_i))^{-1}.$ Notice the fact that if an $l\times l$ matrix $A$ and an $l\times l$ invertible matrix $B$ satisfy $AB=BA,$ then $AB^{-1}=B^{-1}BAB^{-1}=B^{-1}ABB^{-1}=B^{-1}A.$ Thus $W_iA_j=A_jW_i$ for any $i,j\in[n].$ 

Denote the codes defined in \eqref{eq:gen} and \eqref{eq:parity} as $\cC_1$ and $\cC_2$ respectively.
By \eqref{eq:identity}, $HG^T=0.$ Thus $\cC_1\subseteq \cC_2.$ Since $G$ and $H$ both have full rank, $\cC_1$ and $\cC_2$ have the same dimension $kl$ as a vector space over $F.$ Consequently $\cC_1=\cC_2.$ This completes the proof.
\end{IEEEproof}

\subsection{A family of MDS array codes with the $d$-optimal repair property}
In this section we use the results about GRSA codes to construct a family of $(n,k=n-r,l)$ MDS array codes with the $d$-optimal repair property.

\begin{cnstr}\label{constr7} 
Let $F$ be a finite field of size $|F|\ge n+1$ and let $\gamma$ be a primitive element in $F.$ Let $s=d+1-k$ and $l=s^{n-1}.$
Consider the code family given by \eqref{eq:parityform}-\eqref{eq:power}, where
the matrices $A_1,A_2,\dots,A_n$ are given by
\begin{align*}
A_i=\sum_{a=0}^{l-1}\lambda_{i,a_i}e_a &e_{a(i,a_i\oplus 1)}^T,\; i=1,2,...,n-1,\\
&A_n=I,
\end{align*}
where $\oplus$ denotes addition modulo $s, \lambda_{i,0}=\gamma^i$ for all $i\in[n-1]$ and $\lambda_{i,u}=1$ for all $i\in[n-1]$ and all $u\in\{1,2,\dots,s-1\}.$ Here
$\{e_a:a=0,1,\dots,l-1\}$ is the standard basis of $F^l$ over $F,$ $a_i$ is the $i$-th digit from the right in the
representation of $a=(a_{n-1},a_{n-2},
\dots,a_1)$  in the $s$-ary form, and $a(i,u)$ is defined in the same way as in Sect.~\ref{opt}.
\end{cnstr}
\begin{theorem}
The code $\cC$ given by Construction \ref{constr7} is an MDS array code.
\end{theorem}
\begin{IEEEproof} 
The proof parallels the proof of Theorem \ref{3MDS}: we show that  for any $i\neq j,$ $ A_iA_j=A_jA_i$ and that the matrix $A_i-A_j$ is invertible. Thus, $\cC$ is an MDS array code.
\end{IEEEproof}
\begin{theorem}\label{cons4}
The code $\cC$ given by Construction \ref{constr7} has the $d$-optimal repair property.
\end{theorem}
\begin{IEEEproof}
By Theorem \ref{dualgrsa}, $\cC=\text{\rm GRSA}(n,k,\sA,\sV)$ with $\sA=\{A_1,A_2,\dots,A_n\}$ and $\sV=\{V_1,\dots,V_n\},$ where $V_i=(\prod_{j\neq i}(A_j-A_i))^{-1}.$ Suppose that we want to use $C_{i_2},C_{i_3},\dots,C_{i_{d+1}}$ to repair $C_{i_1}.$ Let $\Delta=\{i_1,i_2,\dots,i_{d+1}\}.$ Clearly
$(C_{i_1},C_{i_2},\dots,C_{i_{d+1}})$ is a Generalized Reed-Solomon Array code $\text{\rm GRSA}(d+1,k,\sA_{\Delta},\sV_{\Delta}),$ where  $\sA_{\Delta}=\{A_{i_1},A_{i_2},\dots,A_{i_{d+1}}\}$ and $\sV_{\Delta}=\{V_{i_1},V_{i_2},\dots,V_{i_{d+1}}\}.$ By Theorem \ref{dualgrsa}, there is a set of $l\times l$ invertible matrices $\sW_{\Delta}=\{W_1,W_2,\dots,W_{d+1}\}$ such that
\begin{equation}\label{eq:wc}
\left[ \begin{array}{cccc} I & I & \dots & I\\
A_{i_1} & A_{i_2} & \dots & A_{i_{d+1}}\\
A_{i_1}^2 & A_{i_2}^2 & \dots & A_{i_{d+1}}^2\\
\vdots & \vdots & \vdots & \vdots\\
A_{i_1}^{s-1} & A_{i_2}^{s-1} & \dots & A_{i_{d+1}}^{s-1} \end{array} \right]
\left[ \begin{array}{c} W_1C_{i_1}\\
W_2C_{i_2}\\
\vdots \\
W_{d+1}C_{i_{d+1}} \end{array} \right]=0,
\end{equation}
Using the same method as in the proof of Theorem \ref{opacs}, we can show that $W_1C_{i_1}$ can be recovered by downloading a vector in $F^{l/s}$ from each of the nodes $C_{i_2},C_{i_3},\dots,C_{i_{d+1}}.$ Since $W_1$ is invertible, $C_{i_1}$ can be recovered by the same set of vectors. This shows that $\cC$ has the $d$-optimal repair property.
\end{IEEEproof}
\subsection{Extension to $d$-optimal repair property for several values of $d$ simultaneously}
Now we give a simple extension of the previous construction to make the code have $d$-optimal repair property for several values of $d$ simultaneously. More specifically, give any positive integers $n,k,m,d_1,d_2,\dots,d_m$ such that $k\le d_1,\dots,d_m<n,$ we will show that by replacing $s$ in Construction \ref{constr7} with the value
$$
s=\text{lcm}(d_1+1-k,d_2+1-k,\dots,d_m+1-k),
$$
we will obtain an $(n,k,l=s^{n-1})$ MDS array code $\cC$ with $d_i$-optimal repair property for all $i=1,\dots,m$ simultaneously.

By the proof of Theorem \ref{3MDS}, we know that $\cC$ is an MDS array code. Now we prove that $\cC$ has $d_i$-optimal repair property for any $i\in[m].$

\begin{theorem}
The code $\cC$ has $d_i$-optimal repair property for any $i\in[m].$
\end{theorem}
\begin{IEEEproof}
Given any $i\in[m],$ we show that $\cC$ has $d_i$-optimal repair property. Let $s_i=d_i+1-k.$ Without loss of generality, we only prove the case when we use $C_2,C_3,\dots,C_{d_i+1}$ to repair $C_1.$ (The proof below needs some slight modifications for the case of repairing $C_n,$ we omit this special case here.) Following exactly the same steps in the proof of Theorem \ref{cons4}, we obtain
\begin{equation}\label{eq:newwc}
\left[ \begin{array}{cccc} I & I & \dots & I\\
A_1 & A_2 & \dots & A_{d_i+1}\\
A_1^2 & A_2^2 & \dots & A_{d_i+1}^2\\
\vdots & \vdots & \vdots & \vdots\\
A_1^{s_i-1} & A_2^{s_i-1} & \dots & A_{d_i+1}^{s_i-1} \end{array} \right]
\left[ \begin{array}{c} W_1C_1\\
W_2C_2\\
\vdots \\
W_{d_i+1}C_{d_i+1} \end{array} \right]=0,
\end{equation}
where $W_1,W_2,\dots,W_{d_i+1}$ are some $l\times l$ invertible matrices. Let $C_i'=W_iC_i.$ Let $c'_{i,a}$ denote the $a$-th coordinate of the column vector $C'_i$ for all $a=0,\dots,l-1.$ We can write out the parity-check equations \eqref{eq:newwc} coordinatewise:
\begin{equation}\label{eq:epp}
\begin{aligned}
&\sum_{i=1}^{d_i+1} \beta_{i,a_i,t} c'_{i,a(i,a_i\oplus t)}=0\\
\text{ for all } t=0,&1,\dots,s_i-1 \text{ and } a=0,\dots,l-1,
\end{aligned}
\end{equation}
 where $\oplus$ denotes addition modulo $s,$ $\beta_{i,u,0}=1$ and $\beta_{i,u,t}=\prod_{v=u}^{u\oplus (t- 1)}\lambda_{i,v}$ for $t=1,\dots,s_i-1.$
Clearly \eqref{eq:epp} indicates that for any $t=0,1,\dots,s-1,$ the coordinates $\{c'_{1,a}:a_i=t,t\oplus 1,t\oplus 2\dots,t\oplus (s_i-1)\}$ can be determined by $\{c'_{j,a}:j=2,3,\dots,d_i+1,a_i=t\}.$  Thus $C'_1$ can be determined by $\{c_{j,a}:j=2,3,\dots,d_i+1,a_i=0,s_i,2s_i,3s_i,\dots,s-s_i\}.$ Since $s_i|s,$ we conclude that $C'_1$ and thus $C_1$ itself can be recovered by downloading a vector in $F^{l/s_i}$ from each of the nodes $C_2,C_3,\dots,C_{d_i+1}.$ This completes the proof.
\end{IEEEproof}

\begin{corollary}
The $(n,k,(n-k)^{n-1})$ MDS array code given by Construction \ref{constr4} has $d$-optimal repair property if $(d+1-k)|(n-k).$
\end{corollary}

\section{Conclusion}
The main problems related to the construction of high-rate regenerating codes are concerned with finding explicit (non-probabilistic)
code families over a field $F$ of small size, with a small subpacketization value $l,$ optimal access or repair bandwidth, and
optimal error resilience. In this paper we present constructions of such codes over fields of size proportional to the code
length $n$. The remaining open problems are constructing codes with similar properties over fields of constant size (independent of $n$) and with smaller values of $l$.

\section*{Acknowledgment} 
We are grateful to our colleague Itzhak Tamo for his help with the proof of Theorem \ref{help}.

\bibliographystyle{IEEEtran}
\bibliography{optimal}

\end{document}